\documentclass[10pt,twocolumn,aps,tightenlines,floats,amsmath,amssymb,prd,floatfix,longbibliography,superscriptaddress,notitlepage,nofootinbib]{revtex4-2}

\usepackage[utf8]{inputenc}
\usepackage[T1]{fontenc}
\usepackage{wasysym}
\usepackage{amsmath,amssymb,amsfonts,amsthm,mathrsfs}
\usepackage{graphicx,wrapfig}
\usepackage{enumerate}
\usepackage{enumitem}
\usepackage[table]{xcolor}
\definecolor{lightblue}{RGB}{51,102,204} 
\definecolor{lightred}{RGB}{191,60,45} 
\usepackage{physics}
\usepackage{mathtools}
\usepackage{dcolumn}
\usepackage{orcidlink}
\usepackage{bbm}
\usepackage{xcolor}
\usepackage{tikz}
\usetikzlibrary{calc}
\usetikzlibrary{arrows.meta}

\usepackage{hyperref}
\hypersetup{
	bookmarks=true,         
	unicode=false,          
	pdftoolbar=true,        
	pdfmenubar=true,        
	pdffitwindow=false,     
	pdfstartview={FitH},    
	pdftitle={Toller matrices and the Feynman i\varepsilon in spinfoams},    
	pdfauthor={ Eugenio Bianchi, Chaosong Chen,Mauricio Gamonal},     
	pdfkeywords={causality, loop quantum gravity, quantum gravity}, 
	pdfnewwindow=true,      
	colorlinks=true,       
	linkcolor=lightred,          
	citecolor=lightblue,        
	filecolor=cyan,         
	urlcolor=lightblue        
}

\newcommand{\ii}{\mathrm{i}}
\newcommand{\ee}{\mathrm{e}}
\newcommand{\CC}{\mathbb{C}}

\newcommand{\outcomment}[1]{}

\newcommand{\comment}[1]{{\color{purple}#1}}

\begin{document}

\title{\Large Toller matrices and the Feynman \boldmath $\mathrm{i}\varepsilon$ in spinfoams}

\author{Eugenio Bianchi\,\orcidlink{0000-0001-7847-9929}}
\email{ebianchi@psu.edu}
\author{Chaosong Chen\,\orcidlink{0009-0003-6722-2126}}
\email{cchen@psu.edu}
\author{Mauricio Gamonal\,\orcidlink{0000-0002-0677-4926}}
\email{mgamonal@psu.edu}

\affiliation{Institute for Gravitation and the Cosmos, The Pennsylvania State University, University Park, Pennsylvania 16802, USA}\affiliation{Department of Physics, The Pennsylvania State University, University Park, Pennsylvania 16802, USA}

\date{\today}

\begin{abstract}
 We study the analytic properties and three equivalent representations of the Toller matrices $T^{(\pm)}$ which appear in the causal formulation of spinfoam transition amplitudes for 4d Lorentzian quantum gravity. These are polynomially bounded functions on the Lorentz group which satisfy the relation $T^{(+)}+T^{(-)}=D$, where the Wigner matrix $D$ provides a unitary irreducible representation of $SL(2,\mathbb{C})$. R\"uhl's definition of $T^{(\pm)}$ in terms of analyticity and asymptotic properties is shown to be equivalent to the recently introduced Feynman $\mathrm{i}\varepsilon$ prescription in spinfoams. We show that, equivalently, they can be represented as an integral over eigenvalues of the boost operator, which results in a sum over residues. The latter reproduces the Wick rotation relating Euclidean $Spin(4)$ to Lorentzian $SL(2,\mathbb{C})$  spinfoams studied by Don\`a, Gozzini and Nicotra. We provide explicit expressions in terms of hypergeometric functions and specialize them to the $\gamma$-simple representations relevant for spinfoams.
\end{abstract}

\maketitle




\section{Introduction}
\label{sec:Introduction}

Harmonic analysis on the Lorentz group \cite{Ruhl:1970lor,Martin-Dussaud:2019ypf} plays a central role in the construction of transition amplitudes in quantum field theory and Lorentzian quantum gravity. In spinfoams, namely the path-integral formulation of loop quantum gravity (LQG) \cite{Rovelli:2014ssa,Ashtekar:2021kfp}, the unitary irreducible representations of $SL(2,\CC)$, which are encoded in Wigner $D$-matrices \cite{Martin-Dussaud:2019ypf}, provide the natural building blocks to construct the Engle–Pereira–Rovelli–Livine (EPRL) vertex amplitude \cite{Engle:2007wy}. By construction, the EPRL model includes an unconstrained sum over combinatorial quantities encoding spinfoam edge orientations.

In the companion paper \cite{Bianchi:2026rjd}, we introduced a new spinfoam vertex amplitude which enforces causality in the EPRL model. The causal vertex is defined by a constrained sum over combinatorial quantities defining a suitable notion of discrete causal structures, determined by the orientation of the edges at a given vertex and its corresponding combinatorial causal data \cite{Livine:2002rh,Oriti:2004mu,Bianchi:2021ric} (see also \cite{Reisenberger:1996pu,Markopoulou:1997wi,Markopoulou:1997hu,Markopoulou:1999cz,Gupta:1999cp,Pfeiffer:2002ic,Hawkins:2003vc,Freidel:2005bb,Oriti:2005jr,Oriti:2006wq,Livine:2006xc,Rovelli:2012yy,Bianchi:2012nk,Oriti:2013aqa,Immirzi:2013rka,Cortes:2014oka,Wieland:2014nka,Immirzi:2016nnz,Finocchiaro:2018hks,Jercher:2022mky,Jercher:2024kig,Simao:2024don,Oriti:2025uad,Asante:2025qbr,Beltran:2026eac}). Remarkably, the building block that encodes these additional constraints is not the Wigner $D$-matrix but a different mathematical object: the \emph{Toller $T$-matrices}. Their analytic properties are governed by the presence of Toller poles, which were studied in the context of relativistic scattering theory \cite{Toller:1968gr,Toller:1968pole,Sciarrino:1967}, and from which we adopt the terminology. Although the Toller $T$-matrices were already introduced by Rühl in the context of the harmonic analysis on the Lorentz group \cite{Ruhl:1970lor}, their direct implementation within the spinfoam formalism has so far remained limited.

In this work, we focus on the study of the analytic properties associated with the Toller $T$-matrices and provide three equivalent representations for them which are directly relevant for applications in spinfoams.

\definecolor{feynmanGreen}{RGB}{222,239,233}
\definecolor{peerBlue}{RGB}{225,237,249}
\definecolor{centerPurple}{RGB}{235,230,244}
\definecolor{armPumpkin}{HTML}{FF7518}
\newcommand{\FillAlpha}{0.6}

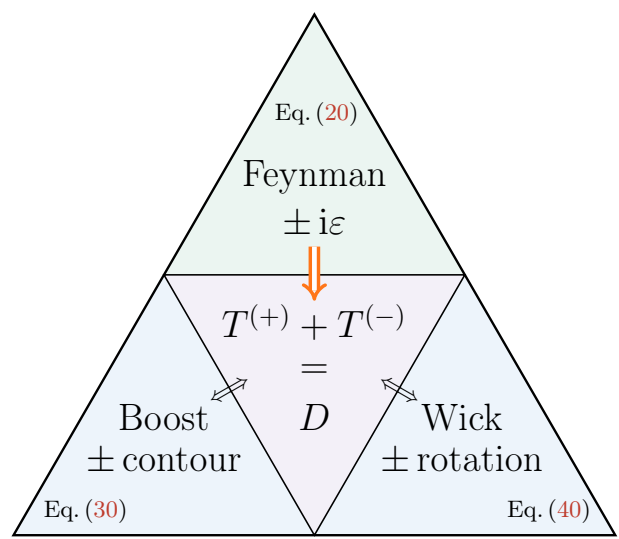
\begin{figure}[!b]
\centering
\begin{tikzpicture}[scale=8]
\coordinate (A) at (0,0);
\coordinate (B) at (1,0);
\coordinate (C) at (0.5,{sqrt(3)/2});

\def\r{0.5}
\coordinate (Ai) at ($(A)!0.5!(B)$);
\coordinate (Bi) at ($(B)!0.5!(C)$);
\coordinate (Ci) at ($(C)!0.5!(A)$);

\begin{scope}[fill opacity=\FillAlpha]
  \fill[feynmanGreen] (Ci) -- (C)  -- (Bi) -- cycle;   
  \fill[peerBlue]     (A)  -- (Ai) -- (Ci) -- cycle;   
  \fill[peerBlue]     (Ai) -- (B)  -- (Bi) -- cycle;   
  \fill[centerPurple] (Ai) -- (Bi) -- (Ci) -- cycle;   
\end{scope}

\draw[line width=0.9pt] (A)--(B)--(C)--cycle;
\draw[line width=0.6pt] (Ai)--(Bi)--(Ci)--cycle;

\coordinate (TopCen) at (0.5,0.56);
\coordinate (BLcen)  at (0.25,0.16);
\coordinate (BRcen)  at (0.745,0.16);
\coordinate (CentCen) at (0.5,0.28);

\node[align=center] at (TopCen) {\Large Feynman\\[5pt]  \Large $\pm \, \mathrm{i}\varepsilon$};
\node[align=center] at (0.504,0.70) {Eq.\,\eqref{eq:ieps-reduced-toller}};

\draw[-{Implies[length=6pt,width=8pt]}, double, double distance=1.8pt,
      line width=1.3pt, line cap=butt, color=armPumpkin]
    (0.5, 0.48) -- (0.5, 0.39);

\node[align=center] at (CentCen) {\Large $T^{(+)} + T^{(-)}$\\[5pt] \Large $=$\\[9pt] \Large $D$};
\node[align=center] at (BLcen) { 
{\Large Boost} \\[5pt]\Large $\pm\,$contour };
\node[align=center] at (0.12,0.04) {Eq.\,\eqref{eq:t-as-boost-contour}};
\node[rotate = 30] at (0.36,0.24) {$\Longleftrightarrow$};
\node[align=center] at (BRcen) {\Large Wick \\[5pt] \Large $\pm\,$rotation  };
\node[align=center] at (0.89,0.04) {Eq.\,\eqref{eq:t-as-Wick-rotation}};
\node[rotate = -30] at (0.64,0.24) {$\Longleftrightarrow$};
\end{tikzpicture}

	\caption{Outline of the paper: Equivalent representations of Toller $T$-matrices and their relation to the Wigner $D$-matrix.
    }
	\label{fig:Triangle}
\end{figure}

First, we show how the mathematical definition of the Toller $T$-matrices in terms of their analytic and asymptotic properties \cite{Ruhl:1970lor} is completely encoded in the novel $\rho$-integral representation introduced in \cite{Bianchi:2026rjd} which determines the Feynman $\ii \varepsilon$ prescription for spinfoams. Then, we show that this formulation is equivalent to an $\omega$-integral representation, where $\omega$ labels eigenvalues of the Lorentz boost along the $z$-axis. Starting from the complexified Lorentz algebra, the reduced Toller matrices emerge as the sum over residues of the overlap coefficients $\bra{\omega\, m}\ket{j\, m}$. Finally, we discuss a third equivalent formulation, where the Toller matrices naturally appear after implementing a notion of Wick rotation between the Euclidean theory described by the group $Spin(4)$ and the Lorentzian theory described by the group $SL(2,\CC)$ \cite{Dona:2021ldn}. We then determine explicit expressions for the reduced Toller matrices in terms of hypergeometric functions and other related properties which are relevant for the spinfoam vertex amplitude. The three different representations and the structure of the paper are summarized in Fig.~\ref{fig:Triangle}.

\section{Wigner and Toller matrices}
\label{sec:Wigner-Toller}

The spinfoam formulation of $4d$ Lorentzian quantum gravity provides a covariant framework for quantum geometry with Lorentz symmetry, described by the group $\mathrm{SO}^{\uparrow}(1,3)$, defining the elementary building blocks. By Wigner's theorem, continuous symmetries are represented in the quantum theory by projective unitary operators, which can be lifted to ordinary unitary representations of the universal (double) cover $\mathrm{SL}(2,\mathbb C)$. A detailed discussion of the representation theory of this group, which plays a central role in the construction of the EPRL model, can be found in \cite{Martin-Dussaud:2019ypf}.  

Here we focus on the unitary, infinite-dimensional irreducible representations of the principal series, $\mathcal{D}^{(\rho,k)}(g)$, which are labeled by a spin $k \in \tfrac{1}{2}\mathbb{Z}$ and a continuous parameter $\rho \in \mathbb{R}$. In the canonical basis of simultaneous eigenstates of $\vec{L}^2$ and $L_z$, denoted by $\ket{(\rho,k);j,m}$, these representations are encoded in the Wigner $D$-matrix, with elements given by
\begin{equation}
D^{(\rho,k)}_{jp\,ln}(g) = \bra{ (\rho,k);j,p} \mathcal{D}^{(\rho,k)}(g)\ket{(\rho,k);l,n}\,.
\end{equation}
The Cartan decomposition allows us to express any group element $g\in SL(2,\CC)$ as 
\begin{equation}
g = U_1 \, \ee^{-\ii\beta K_z} \, U_2\,,
\label{eq:Cartan-SL2C}
\end{equation}
where $U_1,U_2\in SU(2)$, $\beta \ge 0$ is the rapidity of the boost along the axis $z$, and $K_z=\frac{\ii}{2}\sigma_z$ is the $z$-component of the boost generator, with $\sigma_z$ a Pauli matrix. Consequently, the Wigner $D$-matrix can be decomposed as
\begin{equation}
D^{(\rho,k)}_{jp\,ln}(g)=\sum_{m=-\min(j,l)}^{\min(j,l)} D^{(j)}_{pm}(U_1)\,d^{(\rho,k)}_{jlm}(\beta)\,D^{(l)}_{mn}(U_2)\,
\end{equation}
where $D^{(j)}_{mn}(U)$ is the Wigner $D$-matrix of $SU(2)$, and $d^{(\rho,k)}_{jlm}(\beta) \equiv D^{(\rho,k)}_{jmlm} (\ee^{-\ii \beta K_z})$ is the reduced Wigner $d$-matrix, which captures the non-compact part of the representation (see Appendix \ref{app:SL2C} for additional details).

In R\"uhl's monograph \cite{Ruhl:1970lor}, the Wigner $d$-matrices are referred to as ``functions of the first kind'', and are denoted by $d^{(\rho,k)}_{jlm}(\beta)$. These are entire functions in the complex $\rho$-plane, i.e., they have no poles and no branch cuts. 
Throughout this paper we adopt R\"uhl's phase convention, so
that the analytic structure of $d^{(\rho,k)}_{jlm}(\beta)$ is
carried entirely by the Toller $t$-matrices introduced below.
Other conventions commonly used in the spinfoam literature, such as the
one in~\cite{Speziale:2016axj}, differ from R\"uhl's by an
overall $\rho$-dependent phase
\begin{align}
  \Phi(\rho;j,l)
  &= (-1)^{-\frac{j-l}{2}}\,
    \frac{\Gamma(j+\ii\rho+1)}{|\Gamma(j+\ii\rho+1)|}\,
    \frac{\Gamma(l-\ii\rho+1)}{|\Gamma(l-\ii\rho+1)|}\, \nonumber \\[.5em]
    &= (\ee^{-\ii \tfrac{\pi}{2} j} \ee^{+\ii \Psi_{j}^{\rho} } )\; (\ee^{+\ii \tfrac{\pi}{2} l} \ee^{-\ii \Psi_{l}^{\rho}})
  \label{eq:other-phase}
\end{align}
which has unit modulus on the real axis but, once analytically continued,
introduces additional singularities and branch cuts in the complex
$\rho$-plane.

Assuming $\beta>0$, R\"uhl proves the existence and uniqueness of ``functions of the second kind'' that satisfy
\begin{equation}
e^{(\rho,k)}_{jlm}(\beta) = d^{(\rho,k)}_{jlm}(\beta) - (-1)^{j-l}\, e^{(-\rho,-k)}_{ljm}(\beta)  \,.
\end{equation}
Motivated by the role played by Toller poles, studied by M.~Toller in the context of scattering amplitudes \cite{Toller:1968gr,Toller:1968pole,Sciarrino:1967} as one of their defining features, we will refer to these functions of the second kind as the \emph{reduced Toller $t$-matrices}, which we identify as 
\begin{align}
t^{(+,\,\rho,k)}_{jlm}(\beta) &= e^{(\rho,k)}_{jlm}(\beta) \, , \nonumber \\[0.5em]
t^{(-,\,\rho,k)}_{jlm}(\beta) &=(-1)^{j-l}\, e^{(-\rho,-k)}_{ljm}(\beta) \, .
\end{align}
These objects are uniquely characterized by the following asymptotic and analytic properties \cite{Ruhl:1970lor}:
\begin{enumerate}
    \item \textbf{Asymptotic properties}:
    
    Rapid decay across the upper (lower) $\rho$ plane for $t^{(+,\rho,k)}_{jlm}(\beta)$ ($t^{(-,\rho,k)}_{jlm}(\beta)$):
    \begin{equation}
    \label{eq:Toller-Property-1}
        t^{(\pm,\rho,k)}_{jlm}(\beta)=\order{\frac{1}{\abs{\rho}^\alpha}} \, , 
    \end{equation}
    as $\abs{\rho}\to \infty$, and for all $\alpha$ such that
    \begin{equation}
    \begin{cases}
            \alpha \in \mathbb{R}\, &\mathrm{for} \, \Im{\rho}\gtrless 0\\[0.5em]
            0<\alpha<1 &\mathrm{for} \, \Im{\rho}=0
        \end{cases}    
    \end{equation}
    \item \textbf{Matching properties}:

    Matching with $d^{(\rho,k)}_{jlm}(\beta)$ across the lower (upper) $\rho$ plane for $t^{(+,\rho,k)}_{jlm}(\beta)$ ($t^{(-,\rho,k)}_{jlm}(\beta)$):
    \begin{equation}
    \label{eq:Toller-Property-2}
        t^{(\pm,\rho,k)}_{jlm}(\beta)= d^{(\rho,k)}_{jlm}(\beta)+\order{\frac{1}{\abs{\rho}^\alpha}} \, , 
    \end{equation}
    as $\abs{\rho}\to \infty$, and for all $\alpha$ such that
    \begin{equation}
    \begin{cases}
            \alpha \in \mathbb{R}\, &\mathrm{for} \, \Im{\rho}\lessgtr 0\\[0.5em]
            0<\alpha<1 &\mathrm{for} \, \Im{\rho}=0
        \end{cases}
    \end{equation}
    \item \textbf{Pole structure}:

    The matrices are meromorphic in the $\rho$ plane, with a finite number of simple poles (Toller poles) located on the following points:
    \begin{equation}
    \label{eq:Toller-Property-3}
    \ii\, \rho =  -j  \,, (-j+1)\, \,, \ldots \, , (l-1) \, ,  l\, .
    \end{equation}
\end{enumerate}
Consequently, the two reduced Toller $t$-matrices sum up to the reduced Wigner $d$-matrix, as 
\begin{equation}
\label{eq:additive-property-reduced}
t^{(+,\rho,k)}_{jlm}(\beta)+t^{(-,\rho,k)}_{jlm}(\beta) = d^{(\rho,k)}_{jlm}(\beta)\, .
\end{equation}

In Sec.~\ref{sec:Feynman-ieps} we will show that the Feynman
$\ii\varepsilon$ prescription extracts each Toller branch directly
from $d^{(\rho,k)}_{jlm}(\beta)$, and that the analytic properties
\eqref{eq:Toller-Property-1}--\eqref{eq:Toller-Property-3}
together with the sum rule \eqref{eq:additive-property-reduced}
determine $t^{(\pm,\rho,k)}_{jlm}(\beta)$ uniquely.

Using the Cartan decomposition \eqref{eq:Cartan-SL2C}, we can construct analogously the Toller $T$-matrices 
\begin{equation}
T^{(\pm,\rho,k)}_{jp\,ln}(g)=\sum_{m=-\min(j,l)}^{\min(j,l)} D^{(j)}_{pm}(U_1)\,t^{(\pm,\rho,k)}_{jlm}(\beta)\,D^{(l)}_{mn}(U_2)\, .
\end{equation}
We note that, while the Wigner $D$-matrix has a group theoretical interpretation as a unitary irreducible representation of the Lorentz group, the Toller $T$-matrices can be thought of instead as functions over $SL(2,\CC)$ which are polynomially bounded in $\Tr(g g^\dagger)$, specifically $|t^{(\pm,\rho,k)}_{jlm}(\beta)| \sim e^{-(1+|k\pm m|)\beta}$ as $\beta \to \infty$, and fully determined by their analytic properties. Crucially, the Toller matrices do not provide a representation of elements of the Lorentz group, i.e., for $g_1,g_2\in SL(2,\CC)$, we have
\begin{equation}
T^{(\pm,\rho,k)}_{jm\,ln}(g_1 \, g_2) \neq \sum_{l',n'} T^{(\pm,\rho,k)}_{jm\,l'n'}(g_1) \; T^{(\pm,\rho,k)}_{l'n'\, ln}(g_2)  \, .
\end{equation}
Nevertheless, the Toller $T$-matrices satisfy the additive property
\begin{equation}
\label{eq:splitting}
T^{(+,\,\rho,k)}_{jm\,ln}(g)\,+\,T^{(-,\,\rho,k)}_{jm\,ln}(g) = D^{(\rho,k)}_{jm\,ln}(g)\, ,
\end{equation}
and, as discussed in detail in \cite{Ruhl:1970lor},
they play a significant role in the harmonic analysis of the Lorentz group, particularly as a basis for the inverse Fourier transform of distributions over $SL(2,\CC)$. 

\section{Equivalent representations of the Toller matrices}
\label{sec:explicit-representations}

In this section we present three equivalent representations of
the Toller matrices $t^{(\pm,\rho,k)}_{jlm}(\beta)$ defined in
Sec.~\ref{sec:Wigner-Toller}. The first
(Sec.~\ref{sec:Feynman-ieps}) is a Feynman $\ii\varepsilon$
contour integral, which is distinguished in that it extracts each
branch from $d^{(\rho,k)}_{jlm}(\beta)$ directly as an analytic projector, and simultaneously establishes the uniqueness asserted in Sec.~\ref{sec:Wigner-Toller}. The second (Sec.~\ref{sec:boost-residue}) is a residue
sum over the complex eigenvalues of the boost generator $K_z$.
The third (Sec.~\ref{sec:Wick-rotation}) is obtained by Wick
rotation of the Clebsch--Gordan coefficients of $Spin(4)$, and
yields the closed-form hypergeometric expressions originally found by R\"uhl~\cite{Ruhl:1970lor}. We summarize these representations
in Fig.~\ref{fig:Triangle}.

\subsection{Toller matrices from the Feynman \boldmath $\ii \varepsilon$}
\label{sec:Feynman-ieps}

\begin{figure}[!b]
	\centering
\includegraphics[width=0.45\textwidth]{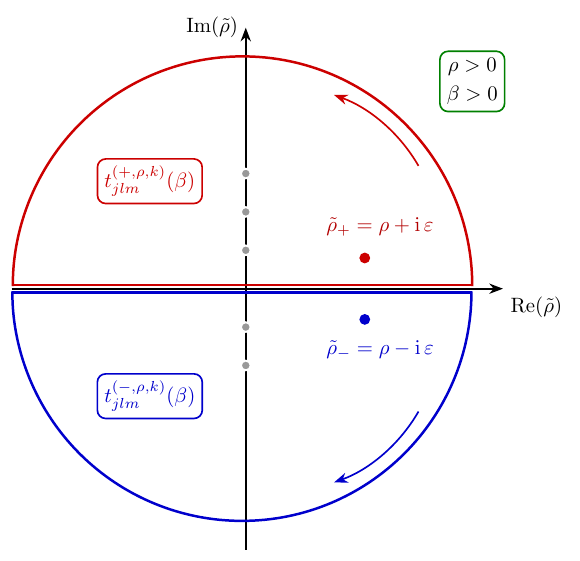}
	\caption{Feynman $\ii\varepsilon$ representation of the reduced
Toller $t$-matrices in \eqref{eq:ieps-reduced-toller}. The contour is
closed in the upper half-plane for $t^{(+,\,\rho,k)}_{jlm}(\beta)$ and
in the lower half-plane for $t^{(-,\,\rho,k)}_{jlm}(\beta)$. Gray
circles mark the zeros of the kernel $P_{jl}(\tilde\rho;\rho)$, which
cancel the Toller poles in the $\tilde\rho$-plane.}
	\label{fig:Toller-Poles}
\end{figure}

In \cite{Bianchi:2026rjd}, we introduced a Toller
kernel defined by
\begin{align}
P_{jl}(\tilde\rho;\rho)
&\equiv
\prod_{n=0}^{j+l}
\frac{\ii\tilde\rho-(n-j)}{\ii\rho-(n-j)}\, \nonumber \\
&= \frac{\Gamma (-j-\ii \rho) \Gamma (l-\ii \tilde{\rho} +1)}{\Gamma (-j-\ii \tilde{\rho} ) \Gamma (l-\ii \rho+1)}\,,
\end{align}
which encodes the Toller poles \eqref{eq:Toller-Property-3}, and satisfies $P_{jl}(\rho;\rho)=1$ and $P_{jl}(\tilde\rho;\rho)=\mathcal{O}(|\tilde\rho|^{j+l+1})$. The additive property \eqref{eq:additive-property-reduced} motivates the use of the Sokhotski--Plemelj identity $\lim_{\varepsilon\to 0^{+}} (\tfrac{1}{x-\ii\varepsilon}-\tfrac{1}{x+\ii \varepsilon} ) =2\pi \ii\,\delta(x)$. Let us define a functional $I_\varepsilon^{(\pm)}$ acting on
functions $f(\tilde\rho)$ meromorphic in $\tilde\rho$ with poles
confined to the Toller poles \eqref{eq:Toller-Property-3}, so that
the factor $P_{jl}(\tilde\rho;\rho)$ cancels them:
\begin{equation}
I_\varepsilon^{(\pm)}[f]
\equiv
\int_{-\infty}^{+\infty}
\frac{d\tilde\rho}{2\pi \ii}\,
\frac{\pm\,P_{jl}(\tilde\rho;\rho)}
{\tilde\rho-\rho\mp\ii\varepsilon}\,
f(\tilde\rho)\,.
\end{equation}
Motivated by the matching property \eqref{eq:Toller-Property-2}, we consider the action of this functional on reduced Wigner $d$-matrices, namely $I_\varepsilon^{(\pm)}[d]=I_\varepsilon^{(\pm)}[t^{+}]+I_\varepsilon^{(\pm)}[t^{-}]$, where we have omitted other indices and variables for simplicity. We first consider the quantity $I_\varepsilon^{(+)}[t^{+}]$. Since the product $P_{jl}(\tilde{\rho};\rho)\, t^{+}(\tilde{\rho})$ cancels out exactly all the Toller poles in $\tilde{\rho}$, the only remaining pole of the integrand is at $\tilde{\rho} = \rho + \ii \varepsilon$. Using the asymptotic property \eqref{eq:Toller-Property-1}, we can close the contour in the upper half-plane, as shown in Fig.~\ref{fig:Toller-Poles}. Hence, Cauchy's residue theorem gives
\begin{align}
\lim_{\varepsilon\to 0^{+}} I_\varepsilon^{(+)}[t^+]
&= \lim_{\varepsilon\to 0^{+}}
\operatorname*{Res}_{\tilde\rho=\rho+\ii\varepsilon}
\left[
\frac{P_{jl}(\tilde\rho;\rho)}
{\tilde\rho-\rho-\ii\varepsilon}\,
t^+(\tilde\rho)
\right]\nonumber \\
&=\lim_{\varepsilon\to 0^{+}}
P_{jl}(\rho+\ii\varepsilon;\rho)\,
t^+(\rho+\ii\varepsilon) \nonumber\\
&= t^{+}(\rho)\, .
\end{align}
Then we consider the action on the other Toller branch, $I_\varepsilon^{(+)}[t^-]$. Because of \eqref{eq:Toller-Property-1}, we close the contour on the lower half-plane. The product $P_{jl}(\tilde{\rho};\rho)\, t^{-}(\tilde{\rho})$ cancels all the Toller poles again, but in this case the only remaining pole $\tilde{\rho}=\rho + \ii \varepsilon$ is not inside the contour and the integral vanishes: $\lim_{\varepsilon\to 0^{+}} I_\varepsilon^{(+)}[t^-] = 0$. Similarly, we find  $\lim_{\varepsilon\to 0^{+}} I_\varepsilon^{(-)}[t^-] = t^{-}$ and $\lim_{\varepsilon\to 0^{+}} I_\varepsilon^{(-)}[t^+]=0$. Consequently, we find that the Toller $t$-matrices can be expressed as a Feynman $\ii\varepsilon$ prescription applied to the reduced Wigner $d$-matrices, as 
\begin{equation}
t^{\pm} = \lim_{\varepsilon\to 0^{+}} I_\varepsilon^{(\pm)}[d]\, ,
\end{equation}
or more explicitly, we can write:
\onecolumngrid
\begin{equation}
\setlength{\fboxsep}{9pt}
\boxed{
t^{(\pm,\,\rho,k)}_{jlm}(\beta)=\lim_{\varepsilon \to 0^+}\int_{-\infty}^{+\infty}\frac{\dd{\tilde{\rho}}}{2\pi\ii}\;\frac{\pm1}{\tilde{\rho}-\rho\mp\ii\, \varepsilon}\;\qty(\prod_{n=0}^{j+l}\frac{\ii\, \tilde{\rho}- \,(n-j)}{\ii\,\rho-(n-j)} )\;d^{(\tilde{\rho},k)}_{jlm}(\beta)\,.
\label{eq:ieps-reduced-toller}
}
\end{equation}
\twocolumngrid
This integral representation was first presented in the companion
paper~\cite{Bianchi:2026rjd}, where it defines the elementary building block of causal spinfoams; the present derivation is
self-contained and makes the additive property
\eqref{eq:additive-property-reduced} manifest.

\textbf{Uniqueness.} The same contour argument establishes that the
Toller splitting \eqref{eq:splitting} is unique. Let $u_\pm(\rho,\beta)$ be functions
meromorphic in $\rho$ satisfying the asymptotic decay
\eqref{eq:Toller-Property-1}, the pole structure
\eqref{eq:Toller-Property-3}, and the sum rule
$u_++u_-=d^{(\rho,k)}_{jlm}(\beta)$. Applying $I^{(\pm)}_\varepsilon$ to
the sum rule and repeating the residue calculation above with $u_\pm$
in place of $t^{(\pm)}$ gives
$u_\pm=\lim_{\varepsilon\to 0^{+}}I^{(\pm)}_\varepsilon[d]
=t^{(\pm)}$. The Feynman $\ii\varepsilon$
prescription therefore acts as a projector onto each admissible Toller
branch.

The choice of half-plane in the prescription
\eqref{eq:ieps-reduced-toller} is not arbitrary: as we will show in
Sec.~\ref{sec:Wick-rotation}, the reduced Wigner $d$-matrix admits a
decomposition into two boost-frequency components in $\rho$,
proportional to $\ee^{\pm\ii\beta\rho}$, and the contours in the upper
and lower half-planes project onto these two components. The Toller
matrices $t^{(\pm,\rho,k)}_{jlm}(\beta)$ are therefore the positive-
and negative-frequency parts of $d^{(\rho,k)}_{jlm}(\beta)$ in the
boost parameter, in direct analogy with the decomposition of a
Feynman propagator in ordinary quantum field theory.

Using \eqref{eq:ieps-reduced-toller}, various properties of the Toller $t$-matrices follow immediately from the properties of the reduced Wigner $d$-matrices (see \eqref{eq:app-Properties-Wigner} and \cite{Martin-Dussaud:2019ypf}). In particular, we identify the following relations:
\begin{align}
\overline{t^{(\pm,\,\rho,k)}_{jlm}(\beta)}\;\, &=\, t^{(\pm,\,-\rho,k)}_{jl m}(\beta) \label{eq:t-prop1}\\
\overline{t^{(\pm,\,\rho,k)}_{jlm}(\beta)}\;\, &=\, (-1)^{j-l} \, t^{(\mp,\,\rho,k)}_{lj -m}(\beta) \\
t^{(\pm,\,\rho,k)}_{jlm}(\beta) \;\,&=\, (-1)^{j-l} \, t^{(\mp,\,-\rho,k)}_{lj-m}(\beta)  \\
t^{(\pm,\,\rho,-k)}_{jlm}(\beta) &=\,t^{(\pm,\,\rho,k)}_{jl-m}(\beta) \label{eq:t-prop4}
\end{align}
where the overline indicates complex conjugation. These relations imply that, under the unitary equivalence between the representations $(\rho,k)$ and $(-\rho,-k)$, there is a mapping between the two reduced Toller $t$-matrices,
\begin{equation}
t^{(\pm, -\rho,-k)}_{jlm}(\beta) = (-1)^{j-l}\, t^{(\mp,\,\rho,k)}_{ljm}(\beta) \, .
\end{equation}



\subsection{Toller matrices from the boost $K_z$}
\label{sec:boost-residue}

\begin{figure}[!b]
	\centering
	\includegraphics[width=0.5\textwidth]{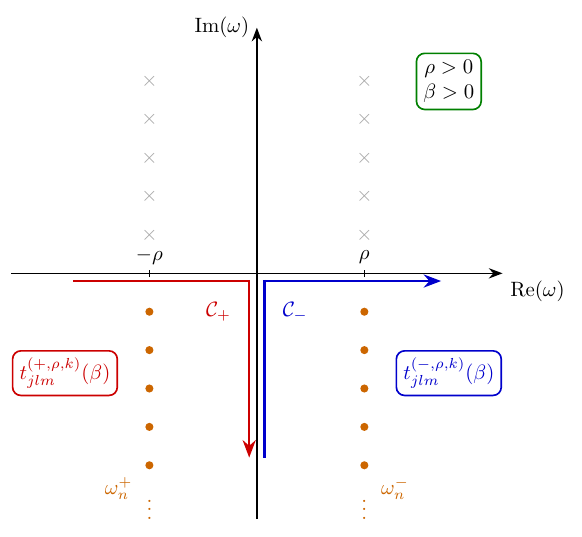}
	\caption{Analytic structure of the integrand in \eqref{eq:d-as-overlaps-Lorentzian}, in the complex $\omega$-plane, where the four families of poles are indicated with crosses and dots. For $\beta>0$, we close the lower-half-plane, where only the poles in \eqref{eq:poles-Kz} (orange dots) are enclosed by the deformation of the contour towards $\omega \to -\ii \infty$. This allows us to express each of the reduced Toller $t$-matrices as a sum over their corresponding residues.}
	\label{fig:Kz-Poles}
\end{figure}

Another convenient representation of the reduced Toller $t$-matrices can be found by writing the matrix elements of the reduced Wigner $d$-matrix in a different basis. We consider the basis built from the simultaneous eigenstates of the boost $K_z$ and the rotation $L_z$, denoted by $\ket{\omega,m}$, such that 
\begin{subequations}
\begin{align}
K_z \ket{ \omega\,m} &= \omega \, \ket{\omega\,m}\,, \\[.5em]
L_z \ket{\omega \,m} &= m \,\ket{\omega\,m}\,,
\end{align}
\end{subequations}
where $\omega \in \mathbb{R}$ and $m\in \tfrac{1}{2}\mathbb{Z}$. The complete set of states $\ket{\omega\, m}$ is orthonormal, $\bra{\omega\,m}\ket{\omega'\, m'} = \delta_{m m'} \; \delta(\omega - \omega')$, and satisfy the resolution of the identity,
\begin{equation}
\label{eq:identity-resolution-booster}
	\mathbbm{1} = \sum_m \int_{-\infty}^{\infty} \dd{\omega} \;  \ket{\omega\, m} \bra{\omega\,m} \, .
\end{equation}
The bases $\ket{\omega\,m}$ and $\ket{j\,m}$ are related by the \emph{overlap coefficients} $\bra{\omega\, m}\ket{j\,m}$, which were computed by Husz\'ar in \cite{Huszar:1971nn} (See also \cite{Rashid:1979xv}), and are reported explicitly in \eqref{eq:Overlap-Definition}. By inserting the resolution of the identity \eqref{eq:identity-resolution-booster} into the definition of the reduced Wigner $d$-matrix, we can write
\begin{align}
d_{jlm}^{(\rho,k)}(\beta) &= D^{(\rho,k)}_{jmlm}(\ee^{\tfrac{\beta \sigma_z}{2}}) \nonumber \\
&=\bra{j\,m}  \ee^{-\ii \beta \hat{K}_z}\ket{l\,m} \nonumber\\
&= \int_{-\infty}^{\infty} \dd{\omega}   \bra{j\,m} \ee^{-\ii \beta K_z} \ket{ \omega\, m } \bra{\omega\, m} \ket{l\,m} \nonumber \\
&= \int_{-\infty}^{\infty} \dd{\omega} \ee^{-\ii \beta \omega} \overline{\bra{ \omega\, m }\ket{j\, m}} \bra{ \omega\, m}\ket{l\, m} \, .
\label{eq:d-as-overlaps-Lorentzian}
\end{align}
The overlap coefficients $\bra{\omega\, m }\ket{j\,m}$ have four families of simple poles in the complex $\omega$ plane. Since the asymptotic behavior of the integrand is $\ee^{-\ii \beta \omega} \ee^{-\pi \abs{\omega}}$ for $\abs{\omega} \to \infty$, we can close the integration contour in the lower-half complex $\omega$ plane for $\beta>0$. Now, if we deform the integration region of \eqref{eq:d-as-overlaps-Lorentzian} into two contours $\mathcal{C}_{+}$ and $\mathcal{C}_{-}$, as shown in Fig.~\ref{fig:Kz-Poles}, each of these contours will enclose a corresponding family of poles $\omega_n^{+}$ and $\omega_n^{-}$:
\begin{equation}
    \label{eq:poles-Kz}
\omega_n^{\pm}  = \mp\rho \, -\,  \ii \, \big( 2n+\abs{k\pm m}+1\big) \,,
\end{equation}
with $n\in \mathbb{N}_0$. These poles are found by analysing the explicit expressions for the overlap coefficients, as discussed in Appendix \ref{app:self-dual}. Using Cauchy's residue theorem, the integral can then be expressed as a sum over the residues of the relevant poles, and we find that the reduced Toller $t$-matrices can be expressed as a contour integral, or more explicitly, as the infinite sum over the associated residues that lie within each contour: 
\onecolumngrid
\begin{equation}
\label{eq:t-as-boost-contour}
\setlength{\fboxsep}{9pt}
\boxed{
t^{(\pm,\,\rho,k)}_{jlm} (\beta) = 
\int_{\mathcal{C}_{\pm}} \dd{\omega} \ee^{-\ii \beta \omega} \overline{\bra{ \omega\, m }\ket{j\, m}} \bra{ \omega\, m}\ket{l\, m} =- 2\pi \ii \, 
\, \sum_{n=0}^{\infty} \ee^{-\ii \beta \omega_{n}^{\pm}}\, \operatorname*{Res}_{\omega =  \omega_n^{\pm}} \qty[ \overline{\bra{\omega\, m} \ket{j\,m} } \bra{\omega\, m} \ket{l\,m}] \, .
}
\end{equation}
\twocolumngrid

One can check by direct computation that the information regarding the location of the Toller poles \eqref{eq:Toller-Property-3} is captured in the residues of the overlap coefficients, while the asymptotic property \eqref{eq:Toller-Property-1} is encoded in the exponential term $\ee^{-\ii \beta \omega_n^{\pm}}$.  Since the additive property \eqref{eq:additive-property-reduced} holds by construction, the matching property \eqref{eq:Toller-Property-2} is consequently satisfied. In this way, the boost $K_z$ representation of the Toller $t$-matrices encodes the defining properties previously identified.

\subsection{Toller matrices from Wick rotation}
\label{sec:Wick-rotation}

As we have described above, both the Feynman $\ii\varepsilon$ formula \eqref{eq:ieps-reduced-toller} and the boost representation \eqref{eq:t-as-boost-contour} provide explicit expressions for the reduced Toller $t$-matrices which use the analytic properties in the complex $\rho$ and $\omega$ plane, respectively. This fact suggests that the properties of the Toller matrices should become manifest within a fully complexified structure of the Lorentz group. Indeed, if one introduces the self-dual and anti-self-dual combinations of the Lorentz generators, $\vec{A}^{\pm} \equiv \frac{1}{2}(\vec{L} \pm \ii \vec{K})$, one finds a decomposition that splits the complexified Lorentz algebra into two commuting $\mathfrak{su}_{\CC}(2)$ algebras, i.e., $\mathfrak{so}_{\CC}(1,3) \cong \mathfrak{su}_{\CC}(2) \oplus \mathfrak{su}_{\CC}(2)$. In this decomposition, the unitary irreducible representations $(\rho,k)$ correspond to pairs of complex spins $(\ell_+,\ell_-)$, defined as
\begin{equation}
\label{eq:complex-spin-Lorentzian}
\ell_+ \equiv \frac{1}{2}(k + \ii \rho -1)\,, \qquad\ell_- \equiv \frac{1}{2}(k - \ii \rho -1)\,.
\end{equation}
With these labels, the new basis $\ket{(\ell_+,\ell_-); \omega, m}$ can be expressed in terms of the simultaneous eigenstates of $A_z^{\pm}$,
\begin{align}
\label{eq:eigenstates-A_pm}
A_z^{\pm} \ket{(\ell_+,\ell_-); \omega, m} &= m_{\pm} \; \ket{(\ell_+,\ell_-); \omega, m}\,, 
\end{align}
where $m_{\pm} = \frac{1}{2}(m \pm \ii\, \omega)$, with $m$ and $\omega$ being the eigenvalues of $L_z$ and $K_z$. In this way, we can define the overlap coefficients as  $\bra{\omega\, m}\ket{j\,m} \equiv \bra{(\ell_+,\ell_-); \omega , m} \ket{(\rho,k);j,m}$. 

With this rich analytic structure in place, we can now explore the relation with the Euclidean theory described in terms of representations of the group $SO(4)$ and the Wick rotation studied by Don\`a, Gozzini and Nicotra in \cite{Dona:2021ldn}. We note that the complexified Lorentz algebra admits a compact real form corresponding to the simply-connected cover of the Euclidean rotation group, $Spin(4)\cong SU(2)\times SU(2)$. Denoting the generators of rotations $\vec{L}$ and Euclidean boosts $\vec{M}$, we can introduce the
self-dual and anti-self-dual combinations $\vec{J}_{L,R}=\tfrac12(\vec{L}\pm\vec{M})$, which generate two commuting $\mathfrak{su}(2)$ subalgebras and realize the canonical decomposition
$\mathfrak{so}(4)\cong \mathfrak{su}(2)_L\oplus \mathfrak{su}(2)_R$. In this decomposition, the
unitary irreducible representations of $Spin(4)$ are labeled by a
pair of spins $(j_L,j_R)\in\tfrac{1}{2}\mathbb{N}_0\times\tfrac{1}{2}\mathbb{N}_0$, that we can parametrize as
\begin{equation}
j_L=\frac{1}{2}(p+q-1), \qquad j_R = \frac{1}{2} (p-q-1)\, ,
\end{equation}
with $p,q \in \tfrac{1}{2}\mathbb{Z}$, which are the Euclidean counterparts of the labels $\rho$ and $k$ of the Lorentzian case. These labels are used to construct a new basis of simultaneous eigenstates of the $z$-component of $\vec{J}_{L,R}$, in analogy to \eqref{eq:eigenstates-A_pm},
\begin{align}
J_{zL}\,| (j_L,j_R); \omega^{(E)} \, m\rangle  &= m_L \, |(j_L,j_R); \omega^{(E)} \, m \rangle \,,\\
J_{zR} \, |(j_L,j_R); \omega^{(E)} \, m\rangle  &= m_R\,  |(j_L,j_R); \omega^{(E)} \, m\rangle \, ,\nonumber 
\end{align}
with $m$ and $\omega^{(E)}$ being the eigenvalues of $L_z$ and $M_z$, respectively, such that $m_L=\tfrac{1}{2} (m+\omega^{(E)})$ and $m_R=\tfrac{1}{2} (m-\omega^{(E)})$. The Cartan decomposition, for $h\in Spin(4)$, reads
$h = u_1 \, \ee^{\ii \, t \, M_z}\, u_2$, where $t \in [0,2\pi) $, and $u_1,\, u_2 \in \mathrm{diag} \, SU(2)$. Consequently, the reduced Wigner $d$-matrix for $Spin(4)$, also known as the Biedenharn-Dolginov function, can be explicitly written by inserting a resolution of the identity in the
$\{\ket{j_L,m_L}\otimes\ket{j_R,m_R}\}$ basis \cite{Biedenharn:1961drs,Barut:1976ub,Lorente:2003wn},
\begin{align}
&d_{jlm}^{(p,q)}(t) = D^{(p,q)}_{jmlm}(\ee^{\ii \, t \, M_z} ) \label{eq:d-as-overlaps-Euclidean}
\\[0.5em]
&\quad= \bra{(p,q);j,m} \ee^{\ii \, t M_z }  \ket{(p,q);l,m} \nonumber \\
&\quad = \sum_{\omega^{(E)}=-j_L-j_R}^{j_L+j_R} \! \ee^{\ii\, t\,   \omega^{(E)}} \;\overline{\langle \omega^{(E)} \, m | j\,m \rangle} \langle \omega^{(E)}\, m |l\, m \rangle\, ,\nonumber 
\end{align}
where the Euclidean overlap coefficients can be written in terms of the usual $SU(2)$ Clebsch-Gordan coefficients,
\begin{align}
\langle \omega^{(E)}\, m | j\,m \rangle &\equiv \langle (j_L,j_R);\omega^{(E)},m | (p,q);j,m \rangle  \\[.5em]
&= C_{j_Lm_L\, j_R m_R}^{jm}\,,\nonumber
\end{align}
with $m=m_L+m_R$. The striking similarity between \eqref{eq:d-as-overlaps-Euclidean} and \eqref{eq:d-as-overlaps-Lorentzian} suggests an underlying relation between the Euclidean and Lorentzian theories for a specific analytic continuation of the labels, i.e., a \emph{Wick rotation}. The role of a Wick rotation in spinfoams was identified recently by Don\`a, Gozzini and Nicotra in \cite{Dona:2021ldn}. Here we uncover its relation to the unique decomposition \eqref{eq:splitting} in Toller matrices.

To see this, first note that the sum in \eqref{eq:d-as-overlaps-Euclidean} is over $\omega^{(E)}=m_L-m_R=2m_L-m$. Writing $m_L=n-j_L$, with $n=0,\ldots,2j_L$, or equivalently, the same finite sum can be parametrized by $m_R$. Writing $m_R=j_R-r$ and $m_L=m-j_R+r$, with $r=0,\ldots,2j_R$, gives the identity
\begin{equation}
    d^{(p,q)}_{jlm}(t)\;=\;d^{(p,q)}_{L\,jlm}(t)\;=\;d^{(p,q)}_{R\,jlm}(t)\,,
\end{equation}
with
\begin{align}
\label{eq:d-as-Clebsch-Gordan}
d^{(p,q)}_{L\,jlm}(t)
\; &=\;
\sum_{n=0}^{2j_L}
\ee^{\ii t\,(2n-2j_L-m)}\;
C^{jm}_{j_L(n-j_L)\,j_R(m-n+j_L)}\nonumber \\[.5em]
&\qquad\quad \times
C^{lm}_{j_L(n-j_L)\,j_R(m-n+j_L)}\,,
\\[1em]
d^{(p,q)}_{R\,jlm}(t)& =
\sum_{r=0}^{2j_R}
\ee^{\ii t\,(m-2j_R+2r)}\;
C^{jm}_{j_L(m-j_R+r)\,j_R(j_R-r)}\nonumber \\[.5em]\label{eq:d-as-Clebsch-Gordan-r}
&\qquad\quad \times
C^{lm}_{j_L(m-j_R+r)\,j_R(j_R-r)}\,.
\end{align}
Following \cite{Dona:2021ldn}, we implement the Wick rotation directly at the level of the coefficient expansions. The analytically continued $SU(2)$ Clebsch--Gordan coefficient is denoted by $\mathfrak C(j_1,m_1,j_2,m_2;j,m)$ and is defined through the Van der Waerden formula \eqref{eq:VanDerWaerden-CG}. The two maps are summarized in Table \ref{tab:Wick-Rotation-EtoL}.

\begin{table}[h]
\caption{Maps that identify the Wick rotation from the Euclidean representations to the two Toller branches of the Lorentzian representations.}
\label{tab:Wick-Rotation-EtoL}
\begin{ruledtabular}
\begin{tabular}{l r}
\rule{0pt}{3ex}
$\mathcal W_+ :$ Euclidean to Toller$_{+}$ & $\mathcal W_- :$ Euclidean to Toller$_{-}$ \\[0.4em]
\hline
\rule{0pt}{3ex}
$(p,q,t)\;\to\;(\ii\rho,\,k,\,\ii\beta)$
&
$(p,q,t)\;\to\;(-\ii\rho,\,-k,\,\ii\beta)$
\\[0.6em]
$(j_L,j_R)\;\to\;(\ell_{+},\,-1-\ell_{-})$
&
$(j_L,j_R)\;\to\;(-1-\ell_{+},\,\ell_{-})$
\\
 $n=s+n_+$ & $r=s+n_-$ \\
$n_+\equiv\max(0,k+m)$ & $n_-\equiv\max(0,k-m)$
\\
\end{tabular}
\end{ruledtabular}
\end{table}

The substitutions of $(j_L,j_R)$ are induced by $j_L=(p+q-1)/2$ and $j_R=(p-q-1)/2$. Applying $\mathcal W_+$ to the $n$-parametrized sum \eqref{eq:d-as-Clebsch-Gordan}, and $\mathcal W_-$ to the $r$-parametrized sum \eqref{eq:d-as-Clebsch-Gordan-r}, gives the Toller series:
\onecolumngrid
\begin{equation}
\setlength{\fboxsep}{9pt}
\boxed{
\label{eq:t-as-Wick-rotation}
\begin{aligned}
d^{(p,q)}_{jlm}(t) = \begin{cases}
d^{(p,q)}_{L\,jlm}(t)
&\underset{\mathcal W_+}{\hookrightarrow}\,\,\,\,
t^{(+,\rho,k)}_{jlm}(\beta)
\equiv
\sum_{s=0}^{\infty}
\ee^{-\beta(1+2s+\abs{k+m}-\ii\rho)}
\mathfrak C^{(+,\rho,k)}_{jlm,s},
\\[1.1em]
d^{(p,q)}_{R\,jlm}(t)
&\underset{\mathcal W_-}{\hookrightarrow}\,\,\,\,
t^{(-,\rho,k)}_{jlm}(\beta)
\equiv
(-1)^{j-l}\sum_{s=0}^{\infty}
\ee^{-\beta(1+2s+\abs{k-m}+\ii\rho)}
\mathfrak C^{(-,\rho,k)}_{jlm,s}.
\end{cases}
\end{aligned}
}
\end{equation}
\twocolumngrid
Here $2n_+-k-m=\abs{k+m}$ and $2n_- -k+m=\abs{k-m}$. With these shifts, the exponential tail is $e^{-2\beta s}$, while the analytically continued Clebsch--Gordan products grow at most polynomially in $s$; hence both Lorentzian series converge for every $\beta>0$. The coefficients are
\begin{widetext}
\begin{align}
\label{eq:Clebsch-Gordan-products-EtoL}
\mathfrak C^{(+,\rho,k)}_{jlm,s}
&\equiv
\qty((-1)^k\,\ee^{+\ii \tfrac{\pi}{2} j} \,\ee^{-\ii \Psi_{j}^{\rho}})\,
\mathfrak C\!\qty(
\ell_+,\,s+n_+-\ell_+;\,
-1-\ell_-,\,m-s-n_++\ell_+;\,
j,m
)\\[.4em]
&\quad\times
\qty((-1)^k\,\ee^{-\ii \tfrac{\pi}{2} l} \,\ee^{+\ii \Psi_{l}^{\rho}})\,
\mathfrak C\!\qty(
\ell_+,\,s+n_+-\ell_+;\,
-1-\ell_-,\,m-s-n_++\ell_+;\,
l,m
),
\nonumber\\[1em]
\mathfrak C^{(-,\rho,k)}_{jlm,s}
&\equiv
\qty((-1)^k\,\ee^{+\ii \tfrac{\pi}{2} j} \,\ee^{-\ii \Psi_{j}^{\rho}})\,
\mathfrak C\!\qty(
-1-\ell_+,\,m-\ell_-+s+n_-;\,
\ell_-,\,\ell_- -s-n_-;\,
j,m
)
\nonumber\\[.4em]
&\quad\times
\qty((-1)^k\,\ee^{-\ii \tfrac{\pi}{2} l} \,\ee^{+\ii \Psi_{l}^{\rho}})\,
\mathfrak C\!\qty(
-1-\ell_+,\,m-\ell_-+s+n_-;\,
\ell_-,\,\ell_- -s-n_-;\,
l,m
).\nonumber
\end{align}
\end{widetext}

The two expressions \eqref{eq:d-as-Clebsch-Gordan} and \eqref{eq:d-as-Clebsch-Gordan-r} are equivalent in the Euclidean, since they are two parametrizations of the same finite Clebsch--Gordan sum. However, after analytic continuation the two coefficient-level expressions are no longer equivalent term by term. The map $\mathcal W_+$ continues the exponential factor to the Lorentzian frequency $e^{+\ii\rho\beta}$, while $\mathcal W_-$ continues it to $e^{-\ii\rho\beta}$. Thus the Wick rotation separates the two boost-frequency signs selected by the Feynman projectors in (Sec.~\ref{sec:Feynman-ieps}), giving the Toller matrices $ t^{(+,\rho,k)}_{jlm}$ and $ t^{(-,\rho,k)}_{jlm}$. Note also that the coefficients $\ee^{-\beta(1+2s+\abs{k\pm m}\mp\ii\rho)}$ in \eqref{eq:t-as-Wick-rotation} are the ones identified in the boost representation \eqref{eq:poles-Kz}.

The construction \eqref{eq:t-as-Wick-rotation} can be inverted. Setting $\rho\to\mp\ii p$, $k \to \pm q$, and $\beta\to -\ii t$ in the two Toller branches, the coefficients $\mathfrak{C}^{(\pm,\rho,k)}_{jlm,s}$ reduce in the Euclidean to ordinary $SU(2)$ Clebsch-Gordan coefficients up to the $s$-independent convention phase, and the infinite Lorentzian series truncate algebraically by the compact $SU(2)$ support conditions, giving,
\begin{equation}
\label{eq:d-as-inverse-Wick-rotation}
t^{(\pm,\rho,k)}_{jlm}(\beta) \underset{\mathcal W_\pm^{-1}}{\hookrightarrow} 
    d^{(p,q)}_{jlm}(t) 
\end{equation}

While the forward rotation splits $d^{(p,q)}_{jlm}(t)$ into two distinct Lorentzian branches, the inverse rotation collapses both branches back onto the same Biedenharn-Dolginov function. Each Toller branch thus independently recovers the full Euclidean Wigner $d$-matrix $d^{(p,q)}_{jlm}(t)$. The explicit formula is reported in Appendix~\ref{app:Wick-rotation}.


It is important to remark that the three expressions for the reduced Toller $t$-matrices---via Feynman $\ii\varepsilon$ \eqref{eq:ieps-reduced-toller}, from the boost representation \eqref{eq:t-as-boost-contour}, and from Wick rotation \eqref{eq:t-as-Wick-rotation}---are mutually equivalent. Concretely, the expressions \eqref{eq:t-as-Wick-rotation}--\eqref{eq:Clebsch-Gordan-products-EtoL} allow us to write the Toller $t$-matrices in terms of hypergeometric functions as done in \cite{Ruhl:1970lor}:
\begin{widetext}
\begin{align}
\label{eq:t-plus-reduced}
	&t_{j l m}^{(+,\rho,k)}(\beta) = \sqrt{(1+2j)(1+2l)}\;
	   \sqrt{\frac{(j-k)!\,(j+k)!\,(l-k)!\,(l+k)!}{(j-m)!\,(j+m)!\,(l-m)!\,(l+m)!}} \nonumber \\
	&\quad \times 
	   \sum_{n_1}
	   \sum_{n_2}
	   \frac{\Gamma(j+k+m-n_1-n_2+i\rho)}{\Gamma(1+j+i\rho)}
	   (-k-m+n_1+n_2)!\;
	   \binom{j-m}{-k-m+n_1}
	   \binom{l-m}{-k-m+n_2}
	   \binom{j+m}{n_1}
	   \binom{l+m}{n_2} \nonumber \\
	&\qquad \times
	   (-1)^{\,j-l+n_1+n_2}\,
	   e^{\,\beta(-1+k+m-2n_2+i\rho)} \;
	   {}_2F_1\!\left(
		 \begin{matrix}
		   1-k-m+n_1+n_2,\; 1+l-i\rho \\
		   1-j-k-m+n_1+n_2-i\rho
		 \end{matrix}
		 ;\, e^{-2\beta}
	   \right)\, ,
\end{align}
with $n_1 = \max(0,m+k),\ldots , \min(j+m,j+k)$ and $n_2 = \max(0,m+k),\ldots , \min(l+m,l+k)$, and
\begin{align}
\label{eq:t-minus-reduced}
	&t_{j l m}^{(-,\rho,k)}(\beta) = \sqrt{(1+2l)(1+2j)}\;
	   \sqrt{\frac{(l+k)!\,(l-k)!\,(j+k)!\,(j-k)!}{(l-m)!\,(l+m)!\,(j-m)!\,(j+m)!}} \nonumber \\
	&\quad \times 
	   \sum_{n_1}
	   \sum_{n_2}
	   \frac{\Gamma(l-k+m-n_1-n_2-i\rho)}{\Gamma(1+l-i\rho)}
	   (k-m+n_1+n_2)!\;
	   \binom{l-m}{k-m+n_1}
	   \binom{j-m}{k-m+n_2}
	   \binom{l+m}{n_1}
	   \binom{j+m}{n_2} \nonumber \\
	&\qquad \times
	   (-1)^{n_1+n_2}\,
	   e^{\,\beta(-1-k+m-2n_2-i\rho)} \;
	   {}_2F_1\!\left(
		 \begin{matrix}
		   1+k-m+n_1+n_2,\; 1+j+i\rho \\
		   1-l+k-m+n_1+n_2+i\rho
		 \end{matrix}
		 ;\, e^{-2\beta}
	   \right)\, ,
\end{align}
\end{widetext}
with $n_1 = \max(0,m-k),\cdots , \min(l+m,l-k)$ and $n_2 = \max(0,m-k),\cdots , \min(j+m,j-k)$.

These closed forms make the boost-frequency split manifest in a concrete
resummation. In \eqref{eq:t-plus-reduced} the oscillatory factor is
$\ee^{+\ii\beta\rho}$, while in \eqref{eq:t-minus-reduced} it is
$\ee^{-\ii\beta\rho}$; the remaining $\beta$-dependence is through real
exponentials and the argument $\ee^{-2\beta}$ of the hypergeometric
function. Thus the hypergeometric representation realizes explicitly the
two branches selected by the Feynman projectors in
Sec.~\ref{sec:Feynman-ieps}.

\begin{table*}[t]
\centering
\setlength{\textfloatsep}{5pt plus 2pt minus 2pt}
\renewcommand{\arraystretch}{2.4} 
\setlength{\tabcolsep}{30pt} 
\caption{Some examples of reduced Wigner and Toller matrices for different values of $k$, $j$, $l$, and $m$.}
\label{tab:functions_rho1}
\begin{tabular}{l l} 
\hline \hline 

\multicolumn{2}{c}{$\boldsymbol{k=\tfrac{1}{2},\ j=\tfrac{1}{2},\ l=\tfrac{1}{2},\ m=+\tfrac{1}{2}}$}\\
\hline
$d^{(\rho,\tfrac{1}{2})}_{\tfrac{1}{2} \tfrac{1}{2} +\tfrac{1}{2}}(\beta)$ &
$\displaystyle \frac{1}{2\sinh(\beta)^2}\,
\frac{(\cosh(\beta)+2\,\ii\,\rho\,\sinh(\beta))\,\ee^{-\ii\,\rho\beta}-\ee^{+\ii\,\rho\beta}}
{(\rho+\tfrac{\ii}{2})(\rho-\tfrac{\ii}{2})}$ \\[0.5em]

$t^{(+,\rho,\tfrac{1}{2})}_{\tfrac{1}{2} \tfrac{1}{2} +\tfrac{1}{2}}(\beta)$ &
$\displaystyle -\frac{\ee^{+\ii\,\rho\beta}}{2\sinh(\beta)^2}\,
\frac{1}{(\rho+\tfrac{\ii}{2})(\rho-\tfrac{\ii}{2})}$\\[0.5em]

$t^{(-,\rho,\tfrac{1}{2})}_{\tfrac{1}{2} \tfrac{1}{2} +\tfrac{1}{2}}(\beta)$ &
$\displaystyle +\frac{\ee^{-\ii\,\rho\beta}}{2\sinh(\beta)^2}\,
\frac{\cosh(\beta)+2\,\ii\,\rho\,\sinh(\beta)}{(\rho+\tfrac{\ii}{2})(\rho-\tfrac{\ii}{2})}$ \\[0.5em]
\hline\hline

\multicolumn{2}{c}{$\boldsymbol{k=\tfrac{1}{2},\ j=\tfrac{1}{2},\ l=\tfrac{1}{2},\ m=-\tfrac{1}{2}}$}\\
\hline
$d^{(\rho,\tfrac{1}{2})}_{\tfrac{1}{2} \tfrac{1}{2} -\tfrac{1}{2}}(\beta)$ &
$\displaystyle \frac{1}{2\sinh(\beta)^2}\,
\frac{(\cosh(\beta)-2\,\ii\,\rho\,\sinh(\beta))\,\ee^{+\ii\,\rho\beta}-\ee^{-\ii\,\rho\beta}}
{(\rho+\tfrac{\ii}{2})(\rho-\tfrac{\ii}{2})}$  \\[0.5em]

$t^{(+,\rho,\tfrac{1}{2})}_{\tfrac{1}{2} \tfrac{1}{2} -\tfrac{1}{2}}(\beta)$ &
$\displaystyle +\frac{\ee^{+\ii\,\rho\beta}}{2\sinh(\beta)^2}\,
\frac{\cosh(\beta)-2\,\ii\,\rho\,\sinh(\beta)}{(\rho+\tfrac{\ii}{2})(\rho-\tfrac{\ii}{2})}$\\[0.5em]

$t^{(-,\rho,\tfrac{1}{2})}_{\tfrac{1}{2} \tfrac{1}{2} -\tfrac{1}{2}}(\beta)$ &
$\displaystyle -\frac{\ee^{-\ii\,\rho\beta}}{2\sinh(\beta)^2}\,
\frac{1}{(\rho+\tfrac{\ii}{2})(\rho-\tfrac{\ii}{2})}$  \\[0.5em]
\hline\hline

\multicolumn{2}{c}{$\boldsymbol{k=1,\ j=1,\ l=1,\ m=0}$}\\
\hline
$d^{(\rho,1)}_{110}(\beta)$ & $\displaystyle \frac{3}{\sinh(\beta)^3}\, \frac{\cosh(\beta)\, \sin(\rho \beta) - \rho \, \sinh(\beta) \, \cos(\rho \beta)}{\rho \, (\rho+\ii) (\rho-\ii)} $\\

$t^{(+,\,\rho,1)}_{110}(\beta)$ & 
$\displaystyle +\frac{3\,\ii\,\ee^{+\ii \rho \beta}}{2\, \sinh(\beta)^3} \, \frac{- \cosh(\beta) + \ii\,\rho\, \sinh(\beta)}
{\rho \, (\rho+\ii) (\rho-\ii)}$ \\

$t^{(-,\,\rho,1)}_{110}(\beta)$ & 
$\displaystyle -\frac{3\,\ii\,\ee^{-\ii \rho \beta}}{2\, \sinh(\beta)^3} \, \frac{- \cosh(\beta) - \ii\,\rho\, \sinh(\beta)}
{\rho \, (\rho+\ii) (\rho-\ii)}$ \\[0.5em]

\hline\hline
\multicolumn{2}{c}{$\boldsymbol{k=1,\ j=1,\ l=1,\ m=-1}$}\\
\hline
$d^{(\rho,1)}_{11-1}(\beta)$ & 
$\displaystyle -\frac{3}{2 \, \sinh(\beta)^3} \, \frac{\sin(\rho\beta) - \rho \,\ee^{\ii \rho \beta} \sinh(\beta) \big(\!\cosh(\beta)- \ii\, \rho \sinh(\beta) \big)}{\rho \, (\rho+\ii) (\rho-\ii)} $ \\

$t^{(+,\,\rho,1)}_{11-1}(\beta)$ & 
$\displaystyle  +\frac{3\,\ii\, \ee^{+\ii \rho\beta}}{4\,\sinh(\beta)^3} \, \frac{(1+\rho^2)-\ii\,\rho\,\big(\!\sinh(2\beta)-\ii\,\rho\cosh(2\beta)\big)}
{\rho \, (\rho+\ii) (\rho-\ii)}$ \\

$t^{(-,\,\rho,1)}_{11-1}(\beta)$ & 
$\displaystyle -\frac{3\,\ii\, \ee^{-\ii \rho\beta}}{4\,\sinh(\beta)^3}  \frac{1}
{\rho \, (\rho+\ii) (\rho-\ii)}$\\[0.5em]

\hline\hline
\multicolumn{2}{c}{$\boldsymbol{k=1,\ j=1,\ l=1,\ m=+1}$}\\
\hline
$d^{(\rho,1)}_{11+1}(\beta)$ & 
$\displaystyle -\frac{3}{2\,\sinh(\beta)^3} \frac{\sin(\beta \rho) - \rho \, \ee^{-\ii \rho \beta} \sinh(\beta)\big(\! \cosh(\beta) + \ii \, \rho \sinh(\beta) \big)}{\rho \, (\rho+\ii) (\rho-\ii)} $  \\

$t^{(+,\,\rho,1)}_{11+1}(\beta)$ & 
$\displaystyle +\frac{3\,\ii\, \ee^{+\ii \rho\beta}}{4\, \sinh(\beta)^3} \, \frac{1}
{\rho \, (\rho+\ii) (\rho-\ii)}$ \\

$t^{(-,\,\rho,1)}_{11+1}(\beta)$ & 
$\displaystyle  -\frac{3\,\ii\, \ee^{-\ii \rho\beta}}{4\,\sinh(\beta)^3} \, \frac{(1+\rho^2)+\ii\,\rho\,\big(\!\sinh(2\beta)+\ii\,\rho\cosh(2\beta)\big)}
{\rho \, (\rho+\ii) (\rho-\ii)}$ \\[0.5em]
\hline\hline
\multicolumn{2}{c}{$\boldsymbol{k=0,\ j=1,\ l=0,\ m=0}$ }\\
\hline
$d^{(\rho,0)}_{100}(\beta)$ & 
$\displaystyle 
-\frac{\ii\sqrt{3}}{\sinh(\beta)^2}\,
\frac{\cosh(\beta)\,\sin(\rho\beta)-\rho\,\sinh(\beta)\,\cos(\rho\beta)}
{\rho\,(\rho-\ii)}$   \\

$t^{(+,\,\rho,0)}_{100}(\beta)$ & 
$\displaystyle 
+\frac{\ii\sqrt{3}\,\ee^{+\ii\rho\beta}}{2\sinh(\beta)^2}\,
\frac{\ii\cosh(\beta)+\rho\,\sinh(\beta)}
{\rho\,(\rho-\ii)}$ \\

$t^{(-,\,\rho,0)}_{100}(\beta)$ & 
$\displaystyle 
+\frac{\ii\sqrt{3}\,\ee^{-\ii\rho\beta}}{2\sinh(\beta)^2}\,
\frac{-\ii\cosh(\beta)+\rho\,\sinh(\beta)}
{\rho\,(\rho-\ii)}$ \\[0.5em]
 \hline \hline 
\end{tabular}
\end{table*}

\section{Application in Spinfoams}
The building blocks of the EPRL spinfoam vertex \cite{Engle:2007wy} and of its causal version \cite{Bianchi:2026rjd} use $\gamma$-simple representations, where $\gamma$ is the Barbero-Immirzi parameter \cite{Rovelli:2014ssa,Ashtekar:2021kfp}. Specifically, one restricts the $d$- and the $t$-matrices to $k=j=l$ and $\rho=\gamma j$. In this case, the Wigner $d$-matrix simplifies considerably, and can be expressed in terms of a single hypergeometric function as \cite{Speziale:2016axj}
\begin{align}
d^{(\rho,j)}_{jjm}(\beta)
&=
\ee^{-(j-\ii\rho+m+1)\beta}\,
{}_2F_1\!\left(
\substack{j+m+1,\; j+1-\ii\rho\\[2pt] 2j+2}
;\,1-\ee^{-2\beta}
\right).
\end{align}
Similarly, the Toller $t$-matrices reduce to a simple expression in terms of a single hypergeometric function \cite{Bianchi:2026rjd}:
\begin{widetext}
\begin{equation}
t^{(\pm,\,\rho,\, j)}_{jjm}(\beta) = \ee^{-(j \,\mp\, \ii \rho \,\pm \,m \,+1)\beta} \; \frac{\Gamma(2j+2)\Gamma(\pm \,\ii \rho \mp m)}{\Gamma(j\,\mp\, m+1)\Gamma(j+1\pm \,\ii \rho)} \; {}_2F_1\!\left(
		 \begin{matrix}
		   j \pm m + 1,\; j+1 \mp \,\ii \rho \\
		    1\pm m \,\mp \,\ii \rho
		 \end{matrix}
		 ;\, \ee^{-2\beta}
	   \right)\,.
\end{equation}
\end{widetext}
As a consistency check, and to illustrate the usefulness of the expressions reported in \eqref{eq:ieps-reduced-toller}, \eqref{eq:t-as-boost-contour}, and \eqref{eq:t-as-Wick-rotation}, we show how to determine the Toller functions explicitly for the Barrett--Crane model \cite{Barrett:1999qw} and its Livine-Oriti causal version \cite{Livine:2002rh, Oriti:2004mu}. This is the case of $\rho\neq 0$ with $k=0$, $j=0$, $l=0$, and $m=0$, which corresponds to the limit $\gamma\to \infty$ at fixed area $\rho=\gamma j$. In this case, the Wigner $d$-matrix is simply
\begin{equation}
d^{(\rho,0)}_{000}(\beta) = \frac{\sin(\rho\, \beta)}{\rho\, \sinh(\beta)}\, .
\end{equation}
The expression for $t^{(+,\rho,0)}_{000}(\beta)$ can be obtained using the $\ii \varepsilon$ prescription \eqref{eq:ieps-reduced-toller}. By closing the upper half-plane we get
\begin{align}
t^{(+,\,\rho,0)}_{000}(\beta) &= \lim_{\varepsilon \to 0^+}\int_{-\infty}^{+\infty}\frac{\dd{\tilde{\rho}}}{2\pi\ii}\;\frac{1}{\tilde{\rho}-\rho-\ii\, \varepsilon} \, \frac{\tilde{\rho}}{\rho} \;\frac{\sin(\tilde{\rho} \beta)}{\tilde{\rho} \sinh(\beta)}  \nonumber \\
&= \frac{1}{2\ii \, \rho \sinh(\beta)}  \,\operatorname*{Res}_{\tilde{\rho}=\rho+\ii \varepsilon}\left[\frac{\ee^{\ii \tilde{\rho} \beta}}{\tilde{\rho}-\rho - \ii\, \varepsilon}\right] \nonumber \\
&= \frac{1}{2\ii\rho} \frac{\ee^{\ii \rho \beta}}{\sinh(\beta)}.
\end{align}
Similarly, for $t^{(-,\rho,0)}_{000}(\beta)$, we close the lower half-plane and find
\begin{align}
t^{(-,\,\rho,0)}_{000}(\beta) &= \lim_{\varepsilon \to 0^+}\int_{-\infty}^{+\infty}\frac{\dd{\tilde{\rho}}}{2\pi\ii}\;\frac{-1}{\tilde{\rho}-\rho+\ii\, \varepsilon} \, \frac{\tilde{\rho}}{\rho} \;\frac{\sin(\tilde{\rho} \beta)}{\tilde{\rho} \sinh(\beta)}  \nonumber \\
&= -\frac{1}{2\ii \, \rho \sinh(\beta)}\lim_{\varepsilon \to 0^+}  \,\operatorname*{Res}_{\tilde{\rho}=\rho-\ii \varepsilon}\left[\frac{\ee^{-\ii \tilde{\rho} \beta}}{\tilde{\rho}-\rho + \ii\, \varepsilon}\right] \nonumber \\
&= -\frac{1}{2\ii\rho} \frac{\ee^{-\ii \rho \beta}}{\sinh(\beta)} \, .
\end{align}
The same result can be obtained using the boost representation \eqref{eq:t-as-boost-contour} with the overlap coefficient reported in \eqref{eq:Overlap-U}, which reads
\begin{equation}
\left|\bra{\omega\,\mu}\ket{0\,0}\right|^2
=
\delta_{\mu,0}\,
\frac{\sinh(\pi \rho)}
{2\rho \big(\cosh(\pi \rho)+\cosh(\pi \omega)\big)}\, .
\end{equation}
The poles \eqref{eq:poles-Kz} relevant to the two Feynman branches here are
\begin{equation}
\omega_n^\pm=-\ii(1+2n)\mp\rho,
\qquad n=0,1,2,\ldots .
\end{equation}
Therefore,
\begin{align}
t^{(\pm,\rho,0)}_{000}(\beta)
&=
-2\pi\ii
\sum_{n=0}^{\infty}
\operatorname*{Res}_{\omega=\omega_n^\pm}
\left[
\frac{\sinh(\pi\rho)\,\ee^{-\ii\beta\omega}}
{2\rho\big(\cosh(\pi\rho)+\cosh(\pi\omega)\big)}
\right]
\nonumber\\
&=
\sum_{n=0}^{\infty}
\left(\pm\frac{1}{\ii\rho}\right)
\ee^{-(1+2n\mp\ii\rho)\beta}
\nonumber\\
&=
\pm\frac{1}{\ii\rho}\,
\frac{\ee^{-(1\mp\ii\rho)\beta}}{1-\ee^{-2\beta}}
\;=\;
\pm \frac{1}{2\ii\rho}\,
\frac{\ee^{\pm\ii\rho\beta}}{\sinh\beta}\, .
\end{align}

Finally, the same result follows from the Wick-rotation representation
\eqref{eq:t-as-Wick-rotation}. For
$k=j=l=m=0$, we have
\begin{equation}
\ell_+ = \frac{\ii\rho-1}{2},
\qquad
\ell_- = \frac{-\ii\rho-1}{2},
\qquad
n_+=n_-=0.
\end{equation}
Using \eqref{eq:Clebsch-Gordan-products-EtoL}, the analytically
continued Clebsch--Gordan products reduce to
\begin{align}
\mathfrak C^{(+,\rho,0)}_{000,s}
&=
\mathfrak C\!\left(
\ell_+,\,s-\ell_+;\,
\ell_+,\,-s+\ell_+;\,
0,0
\right)^2
=
\frac{1}{\ii\rho},
\nonumber\\[.5em]
\mathfrak C^{(-,\rho,0)}_{000,s}
&=
\mathfrak C\!\left(
\ell_-,\,s-\ell_-;\,
\ell_-,\,-s+\ell_-;\,
0,0
\right)^2
=
-\frac{1}{\ii\rho}.
\end{align}
Therefore,
\begin{align}
t^{(\pm,\rho,0)}_{000}(\beta)
&=
\sum_{s=0}^{\infty}
\ee^{-\beta(1+2s\mp\ii\rho)}
\left(\pm\frac{1}{\ii\rho}\right)
\nonumber\\
&=
\pm\frac{1}{\ii\rho}\,
\frac{\ee^{-(1\mp\ii\rho)\beta}}{1-\ee^{-2\beta}}
\;=\;
\pm \frac{1}{2\ii\rho}\,
\frac{\ee^{\pm\ii\rho\beta}}{\sinh\beta}\, .
\end{align}
These expressions show concretely the equivalence of the representations in a simple example. In Table \ref{tab:functions_rho1} we report other examples relevant for the EPRL spinfoam amplitudes \cite{Engle:2007wy} and \cite{Bianchi:2026rjd}, which can be derived in a similar way.


\section{Discussion}
\label{sec:Discussion}
In this paper, we studied the analytic properties of the Toller matrices for the Lorentz group $SL(2,\CC)$, and presented three novel representations, summarized in Fig.~\ref{fig:Triangle}. These equivalent representations uncover different aspects of how the Toller $t$-matrices extract a causal component from the Wigner $d$-matrices: 
\begin{itemize}[leftmargin=1.5em,label={$\triangleright$}]
   \item The Feynman $\pm\ii \varepsilon$ prescription \eqref{eq:ieps-reduced-toller} extracts from the Wigner $d$-matrices the causal components $t^{(\pm,\,\rho,k)}_{jlm}(\beta)$ via a contour integral over the complex variable $\rho\vphantom{t^k}$. This construction encodes exactly, in a single formula, R\"uhl's abstract definition of the $t$ matrices in terms of their pole structure \cite{Ruhl:1970lor}, the Toller poles \cite{Toller:1968gr,Toller:1968pole,Sciarrino:1967}. The other integral representations presented in this paper, together with the properties \eqref{eq:t-prop1}--\eqref{eq:t-prop4} of the Toller $t$-matrices, can be read directly from the expression \eqref{eq:ieps-reduced-toller}. The new expression \eqref{eq:ieps-reduced-toller} was first introduced in \cite{Bianchi:2026rjd} and it plays a central role in the definition of the causal spinfoam vertex for the Lorentzian EPRL model in $4d$. Moreover, the Feynman $\pm\ii \varepsilon$ construction provides an immediate tool for extracting the semiclassical properties of the causal vertex as the contour integral over $\rho$ (Fig.~\ref{fig:Toller-Poles}) selects a single causal spinfoam saddle in the large spin asymptotics \cite{Barrett:2009gg,Barrett:2009mw,Han:2013hna,Dona:2020xzv,Dona:2020yao,Dona:2022hgr,Han:2020fil,Han:2021rjo} (see Sec.~IV in \cite{Bianchi:2026rjd}).

\item The boost representation \eqref{eq:t-as-boost-contour} decomposes the Wigner $d$-matrices into Toller $t$-matrices by splitting the eigenvalues $\omega$ of the boost operator $K_z$ into their positive 
$\mathrm{Re}(\omega)>0$ and negative $\mathrm{Re}(\omega)<0$ parts via the contour integrals $\mathcal{C}_\pm$ (Fig.~\ref{fig:Kz-Poles}). Remarkably, these contour integrals can be written in terms of a sum over residues of simple poles in the complex $\omega$ plane, \eqref{eq:poles-Kz}. This new expression shows clearly that the Toller $t$-matrices are the positive and negative frequency components $\ee^{-\ii \beta \omega_n^\pm}$ of the Wigner $d$-matrices. The frequencies are the poles $\omega_n^\pm$, the time is the boost parameter $\beta$, and the Hamiltonian is the boost operator $K_z$. This structure was first identified in the context of near-horizon entropy in spinfoams \cite{Bianchi:2012ui}, (see also \cite{Bianchi:2012ev,Geiller:2014eza}). Moreover, a defining ingredient of the graviton propagator calculations in spinfoams \cite{Rovelli:2005yj,Bianchi:2006uf,Bianchi:2009ri,Bianchi:2011hp} is the boundary state which determines a semiclassical regime in the bulk. The boundary state has a phase $\sim\ee^{-\ii \gamma j \beta}$ which peaks the boundary geometry on the extrinsic curvature $\beta$ and selects one of the quantum geometries in the bulk. This phenomenon becomes manifest in the boost representation \eqref{eq:t-as-boost-contour} and we expect it will be relevant for the identification of the extrinsic curvature $\beta$ in boundary states for spinfoam cosmology \cite{Vidotto:2010kw,Bianchi:2010zs,Roken:2010vp,Henderson:2010qd,Bianchi:2011ym,Livine:2011up,Rennert:2013qsa,Rennert:2013pfa,Vilensky:2016tnw,Kisielowski:2018oiv,Gozzini:2019nbo,Frisoni:2022urv,Frisoni:2023lvb,Han:2024ydv,Bianchi:2024mrt} and spinfoam black-to-white hole tunneling \cite{Haggard:2014rza,Christodoulou:2016vny,Christodoulou:2018ryl,Bianchi:2018mml,DAmbrosio:2020mut,Soltani:2021zmv,Christodoulou:2023psv,Frisoni:2023agk,Dona:2024rdq,Rovelli:2024sjl,Han:2024rqb,Dona:2025snr}.

\item The Wick rotation formula \eqref{eq:t-as-Wick-rotation} shows that the Lorentzian Toller $t$-matrices can be obtained from the Euclidean Wigner $d$-matrices via a Wick rotation $\mathcal{W}_\pm$ (see Tab.~\ref{tab:Wick-Rotation-EtoL}). The role of a Wick rotation in spinfoams was first identified in \cite{Dona:2021ldn}. Remarkably, the intermediate terms $\mathcal{F}_1$ and $\mathcal{F}_2$ in App.~B of \cite{Dona:2021ldn} (Eq.~112--113) are exactly the Toller $t$-matrices expressed here in terms of the analytically continued Clebsch--Gordan coefficients \eqref{eq:Clebsch-Gordan-products-EtoL}. We note that the new Wick rotation formula \eqref{eq:t-as-Wick-rotation} explains also some of the puzzling behavior found in previous analysis: the Lorentzian Wigner $d$-matrices are obtained only if one sum over the two directions $\mathcal{W}_+$ and $\mathcal{W}_-$ of the Wick rotation, reproducing the decomposition \eqref{eq:splitting}. Moreover the phase $\Psi_j^\rho$ \eqref{eq:other-phase} discussed in \cite{Speziale:2016axj} appears naturally in the Wick rotation from the Euclidean to the Lorentzian, respecting the analyticity properties of the Toller $t$-matrices.
\end{itemize}

\noindent Numerical calculations in spinfoams \cite{Dona:2022yyn}, including the numerical implementation of the EPRL model via the \texttt{sl2cfoam-next} code \cite{Dona:2018nev,Dona:2019dkf,Dona:2020tvv,Gozzini:2021kbt,Dona:2022dxs}, rely on the efficient calculation of Wigner $d$-matrices, for which a decomposition in finite sums of complex exponentials of the form \eqref{eq:t-as-boost-contour} is already being used (see Eq.~6 in \cite{Gozzini:2021kbt}, and \cite{Collet:2018}). We expect that the analytical results presented in this paper will be useful in numerical calculations in spinfoams, including implementations of the causal vertex introduced in \cite{Bianchi:2026rjd}, and computations of the semiclassical boundary state relevant for the graviton propagator \cite{Rovelli:2005yj,Bianchi:2006uf,Bianchi:2009ri,Bianchi:2011hp}, horizon entropy \cite{Bianchi:2012ui,Bianchi:2012ev,Geiller:2014eza}, spinfoam cosmology \cite{Vidotto:2010kw,Bianchi:2010zs,Roken:2010vp,Henderson:2010qd,Bianchi:2011ym,Livine:2011up,Rennert:2013qsa,Rennert:2013pfa,Vilensky:2016tnw,Kisielowski:2018oiv,Gozzini:2019nbo,Frisoni:2022urv,Frisoni:2023lvb,Han:2024ydv,Bianchi:2024mrt} and black-to-white hole tunneling \cite{Haggard:2014rza,Christodoulou:2016vny,Christodoulou:2018ryl,Bianchi:2018mml,DAmbrosio:2020mut,Soltani:2021zmv,Christodoulou:2023psv,Frisoni:2023agk,Dona:2024rdq,Rovelli:2024sjl,Han:2024rqb,Dona:2025snr}.

Self-dual Ashtekar variables, reality conditions, Wick rotations and the map $\gamma\to \pm\ii \gamma$ for the Barbero-Immirzi parameter (see Tab.~\ref{tab:Wick-Rotation-EtoL} with $\rho=\gamma j$), have played a defining role in the early developments of loop quantum gravity \cite{Ashtekar:1986yd,Ashtekar:1987gu,Ashtekar:1991hf,Ashtekar:1991vz,BarberoG:1994eia,Immirzi:1996di,Thiemann:1995ug,Thiemann:1995ug,Ashtekar:1995qw,Rovelli:1997yv} (see also \cite{Wieland:2010ec,Wieland:2011ru,Frodden:2012dq,BenAchour:2014erw,BenAchour:2014qca,Wilson-Ewing:2015lia,BenAchour:2016mnn,BenAchour:2017jof,Varadarajan:2018uaj,Eder:2020okh,Ashtekar:2020xll,Eder:2022gge,Eder:2022eqo,Alexander:2022ocp,Wieland:2023qzp,Sahlmann:2023eqt,Sahlmann:2023xle}). We hope that the results presented in this work will contribute to the future developments of the theory by connecting it to the vast literature on Toller poles for the Lorentz group 
\cite{Toller:1968gr,Toller:1968pole,Sciarrino:1967,Delbourgo:1967ab,Freedman:1967vfr,Dao:1967ri,Bitar:1968lsc,Nakanishi:1968ny,Smorodinsky:1970hv,Smorodinskii:1972kw,Brower:1974yv,Browne:1975ba,Conrady:2010sx}.

\begin{acknowledgments}
We thank Pietro Donà, Abhay Ashtekar, Jonathan Engle, Jeremy Martinon, Muxin Han, Monica Rincon-Ramirez, and Shiji Bi for insightful discussions. M.G. was partially supported by the \href{https://anid.cl}{Agencia Nacional de Investigación y Desarrollo} (ANID) and \href{https://fulbrightchile.cl/}{Fulbright Chile} through the Fulbright Foreign Student Program and ANID BECAS/Doctorado BIO Fulbright-ANID 56190016.~E.B. is supported by the National Science Foundation, Grants No. PHY-2207851 and PHY-2513194. This work was made possible thanks to the support of the WOST project (\href{https://withoutspacetime.org}{\mbox{withoutspacetime.org}}), funded by the John Templeton Foundation (JTF) under Grant ID 63683. 
\end{acknowledgments}

\newpage

\onecolumngrid

\appendix

\section{Unitary irreducible representations of $SL(2,\CC)$}
\label{app:SL2C}

The principal series of unitary irreducible representations of
$SL(2,\CC)$ is infinite-dimensional and is generated by the six
self-adjoint Lorentz generators $J^{IJ}=-J^{JI}$, with
$I,J=0,1,2,3$. Introducing a unit timelike vector $t^I$, we decompose
the generators into rotations that leave $t^I$ invariant,
\begin{equation}
L^I=\frac{1}{2}\epsilon^I{}_{JKL}J^{JK}t^L,
\qquad
K^I=J^{IJ}t_J .
\end{equation}
In coordinates where $t^I=(1,0,0,0)$, we write
$L^I=(0,L^i)$ and $K^I=(0,K^i)$, with $i=1,2,3$. The generators obey
\begin{align}
[L^i,L^j] &= \ii\,\epsilon^{ij}{}_{k}\,L^k,
\nonumber\\
[K^i,K^j] &= -\ii\,\epsilon^{ij}{}_{k}\,L^k,
\nonumber\\
[L^i,K^j] &= \ii\,\epsilon^{ij}{}_{k}\,K^k .
\end{align}
In the defining $2\times2$ matrix realization of the Lorentz algebra,
one may take
\begin{equation}
L^i=\frac{1}{2}\sigma^i,
\qquad
K^i=\frac{\ii}{2}\sigma^i,
\end{equation}
which satisfies the same commutation relations. This finite-dimensional
matrix realization should be distinguished from the Hermitian generators
acting on the unitary principal-series Hilbert space. The principal series of the unitary irreducible representations is labeled by two parameters, $\rho \in \mathbb{R}$ and $k\in \tfrac{1}{2}\mathbb{Z}$, and is infinite-dimensional. Its associated Hilbert space admits the canonical $SU(2)$-decomposition, 
\begin{equation}
\mathcal{H}^{(\rho,k)}=\bigoplus_{j=\abs{k}}^{\infty} \mathcal{H}_j\,,
\end{equation}
where $\mathcal{H}_j$ is the Hilbert space associated to the spin-$j$ representation of $SU(2)$ with dimension $d_j=2j+1$. In the $(\rho,k)$ representation, the invariant Casimir operators have eigenvalues
\begin{subequations}
	\label{eq:Casimirs-SL2C}
\begin{align}
C_1 &= \frac{1}{2} J^{IJ}J_{IJ} = \vec{K}^2 - \vec{L}^2 = \rho^2 - k^2 + 1\, ,\\ C_2 &= \frac{1}{8} \epsilon_{IJKL} J^{IJ} J^{KL} = \vec{K}\cdot \vec{L} = \rho\, k\,.
\end{align}
\end{subequations}
A complete basis of the Hilbert space $\mathcal{H}^{(\rho,k)}$ is given by the orthonormal basis of simultaneous eigenstates of $\vec{L}^2=\delta_{ij}L^i L^j$ and $L_z=L^I z_I$, with $z_I = (0,0,0,1)$, such that
\begin{subequations}
\label{eq:CanonicalBasis-jm}
\begin{align}
C_1\ket{(\rho,k);j,m}&=(\rho^2-k^2+1)\ket{(\rho,k);j,m}\,,\\[.5em]
C_2\ket{(\rho,k);j,m}&=\rho\, k\;\ket{(\rho,k);j,m}\,,\\[.5em]
\vec{L}^2\ket{(\rho,k);j,m}&=j(j+1)\ket{(\rho,k);j,m}\,,\\[.5em]
L_z\ket{(\rho,k);j,m}&=m\;\ket{(\rho,k);j,m}\,,
\end{align}
\end{subequations}
with $\rho\in \mathbb{R}$, $k\in \tfrac{1}{2}\mathbb{Z}$, $\abs{k}\leq j$, and $\abs{m}\leq j$. In the principal series, the unitary representation is a map $\mathcal{D}^{(\rho,k)}(g):SL(2,\CC) \to \mathcal{U}(\mathcal{H}^{(\rho,k)})$, where $\mathcal{U}(\mathcal{H}^{(\rho,k)})$ is the group of all unitary operators on $\mathcal{H}^{(\rho,k)}$, such that $\mathcal{D}^{(\rho,k)}(g^{-1}) = \mathcal{D}^{(\rho,k)}(g)^\dagger$, and
\begin{equation}
	\mathcal{D}^{(\rho,k)}(g_1 g_2) = \mathcal{D}^{(\rho,k)}(g_1)\mathcal{D}^{(\rho,k)}(g_2)\, ,  
\end{equation}
for all $g,g_1,g_2 \in SL(2,\CC)$, and $\ket{\psi} \in \mathcal{H}^{(\rho,k)}$, with $\norm{\mathcal{D}^{(\rho,k)}(g) \ket{\psi}} = \norm{\ket{\psi}}$. The action of the group element $g$ on a state $\ket{\psi}\in \mathcal{H}^{(\rho,k)}$ is given by $\ket{\psi_g}=\mathcal{D}^{(\rho,k)}(g)\ket{\psi}$. In the canonical basis \eqref{eq:CanonicalBasis-jm}, the unitary representations are encoded in the Wigner $D$-matrix, with elements given by
\begin{equation}
D^{(\rho,k)}_{jm\,ln}(g)= \bra{ (\rho,k);j,m} \mathcal{D}^{(\rho,k)}(g)\ket{(\rho,k);l,n} \, ,
\end{equation}
and satisfying the following properties
\begin{align}
D^{(\rho,k)}_{jm\,ln}(g_1 g_2) &= \sum_{l',n'} D^{(\rho,k)}_{jm\,l'n'}(g_1) D^{(\rho,k)}_{l'n'\,ln}(g_2)\,,\\
D^{(\rho,k)}_{jm\,ln}(g^{-1}) &= \overline{D^{(\rho,k)}_{ln\,jm}(g)}\,,\\
\sum_{j,m} \overline{D^{(\rho,k)}_{jm\,l'n'}(g)} D^{(\rho,k)}_{jm\,ln}(g) &= \delta_{ll'}\delta_{nn'}\,,
\end{align}
for all $g,g_1,g_2 \in SL(2,\CC)$. It can be checked that the
representations $(\rho,k)$ and $(-\rho,-k)$ are unitarily equivalent. In
the following we will only need the corresponding identity for the
matrix elements of a pure boost along the $z$-axis.

In the Cartan decomposition, any group element $g\in SL(2,\CC)$ can be
expressed as
\begin{equation}
g = U_1 \, \ee^{-\ii\beta K_z} \, U_2\,,
\label{eq:Cartan}
\end{equation}
where $U_1,U_2\in SU(2)$, $\beta\ge 0$ is the rapidity of the boost
along the $z$-axis, and $K_z=\tfrac{\ii}{2}\sigma_z$ is the
$z$-component of the boost generator. For the pure boost
$g_\beta=\ee^{-\ii\beta K_z}$ the matrix element is diagonal in the
magnetic index,
\begin{equation}
D^{(\rho,k)}_{jm\,ln}(g_\beta)=0
\qquad
\text{unless } m=n .
\end{equation}
With R\"uhl's phase convention, the equivalence between
$(\rho,k)$ and $(-\rho,-k)$ gives
\begin{equation}
D^{(-\rho,-k)}_{jm\,lm}(g_\beta)
=
(-1)^{j-l}\,
D^{(\rho,k)}_{lm\,jm}(g_\beta),
\qquad
m=-\min(j,l),\ldots,\min(j,l).
\label{eq:D-boost-rho-k-equivalence}
\end{equation}
Inserting \eqref{eq:Cartan} into $\mathcal{D}^{(\rho,k)}(g)$ yields the
Cartan decomposition of the Wigner $D$-matrix,
\begin{equation}
D^{(\rho,k)}_{jp\,ln}(g)=\sum_{m=-\min(j,l)}^{\min(j,l)}
D^{(j)}_{pm}(U_1)\,d^{(\rho,k)}_{jlm}(\beta)\,D^{(l)}_{mn}(U_2),
\label{eq:D-Cartan-matrix-elements}
\end{equation}
where $D^{(j)}_{mn}(U)$ is the Wigner $D$-matrix of $SU(2)$, and
$d^{(\rho,k)}_{jlm}(\beta) \equiv D^{(\rho,k)}_{jmlm}(\ee^{-\ii\beta K_z})$
is the reduced Wigner-$d$ function, which encodes the non-compact part of the
representation. An explicit expression as a sum of hypergeometric
functions is available. Following R\"uhl's phase
conventions~\cite{Ruhl:1970lor}, we have
\begin{align}
d^{(\rho,k)}_{jlm}(\beta)
  &= \sqrt{(1+2j)(1+2l)}\,
     \sqrt{\frac{(j-k)!\,(j+k)!\,(-k+l)!\,(k+l)!}
                 {(j-m)!\,(l-m)!\,(j+m)!\,(l+m)!}}
     \;\frac{1}{(1+j+l)!}
     \nonumber\\
  &\quad\times
     \sum_{n_1,n_2} (-1)^{n_1+n_2}
     \binom{j-m}{k-m+n_1}
     \binom{l-m}{k-m+n_2}
     \binom{j+m}{n_1}
     \binom{l+m}{n_2}
     \nonumber\\
  &\quad\times
     (j-k+l+m-n_1-n_2)!\,
     (k-m+n_1+n_2)!\,
     \ee^{\,\beta(-1-k+m-2n_1-\ii\rho)}
     \nonumber\\
  &\quad\times
     {}_{2}F_{1}\Big(
       1+k-m+n_1+n_2,\;
       1+j+\ii\rho;\;
       2+j+l;\;
       1-\ee^{-2\beta}
     \Big)\,,
  \label{eq:d-Ruhl-sum}
\end{align}
where the sums over $n_1,n_2$ run over $n_1 = \max(0,m-k),\ldots,\min(j+m,j-k)$ and $n_2 = \max(0,m-k),\ldots,\min(l+m,l-k)$. With this phase convention, the Wigner $d$-matrices are entire in the complex $\rho$-plane. Note that other phase conventions are typically used in the literature. As we adopted R\"uhl's phase convention to ensure the analyticity of
the Wigner $d$-matrix in the complex $\rho$-plane, our expressions
differ from the ones commonly used in the spinfoam literature, e.g.,
\cite{Speziale:2016axj}, by the overall phase
$\Phi(\rho;j,l)$ defined in \eqref{eq:other-phase}, which has been indicated explicitly within the paper.

With our choice of the overall phase, the reduced Wigner $d$-matrices satisfy the following properties \cite{Martin-Dussaud:2019ypf}:

\begin{subequations}
\label{eq:app-Properties-Wigner}
\begin{align}
\overline{d^{(\rho,k)}_{jlm}(\beta)} &= d^{(-\rho,k)}_{jl m}(\beta) \\
d^{(\rho,k)}_{jlm}(\beta) &= (-1)^{j-l} \, d^{(-\rho,k)}_{lj-m}(\beta)  \\
d^{(\rho,-k)}_{jlm}(\beta) &=d^{(\rho,k)}_{jl-m}(\beta) 
\end{align}
\end{subequations}
which automatically implies the unitary equivalence between the representations $(\rho,k)$ and $(-\rho,-k)$,   
\begin{equation}
(-1)^{j-l}\, d^{(-\rho,-k)}_{ljm}(\beta) = d^{(\rho,k)}_{jlm}(\beta) \, .
\end{equation}


\section{Self-dual representation and eigenvalues of boosts $K_z$}
\label{app:self-dual}

Unitary irreducible representations of $SL(2,\CC)$ can be conveniently analyzed by introducing the self-dual and anti-self-dual combinations of the Lorentz generators, defined as $\vec{A}^{\pm} = \frac{1}{2}(\vec{L} \pm \ii \vec{K})$. These generators provide a decomposition that splits the complexified Lorentz algebra into two commuting $\mathfrak{su}_{\CC}(2)$ algebras, i.e., $\mathfrak{so}_{\CC}(1,3) \cong \mathfrak{su}_{\CC}(2) \oplus \mathfrak{su}_{\CC}(2)$. The self-dual and anti-self-dual generators obey the commutation relations
\begin{equation}
\label{eq:self-dual-Lorentz}
[A_i^{(\pm)},A_j^{(\pm)}]=\ii \, \epsilon_{ij}^{~k} \, A_k^{(\pm)}\,,\qquad\quad	 [A_i^{(+)},A_j^{(-)}]=0\,.
\end{equation}
In this decomposition, the unitary irreducible representations $(\rho,k)$ correspond to pairs of complex spins $(\ell_+,\ell_-)$, defined as
\begin{equation}
\ell_+ = \frac{1}{2}(k + \ii \rho -1)\,, \qquad\quad \ell_- = \frac{1}{2}(k - \ii \rho -1)\,.
\end{equation}
Consequently, the Casimir operators are expressed as
\begin{subequations}
\begin{align}
C_{+} &= (\vec{A}^{+}{})^2 = \ell_{+} (\ell_{+}+1) \\[.5em]
C_{-} &= (\vec{A}^{-}{})^2 =  \ell_{-} (\ell_{-}+1). 
\end{align}
\end{subequations}
The basis states $\ket{(\rho,k);j,m}$ can be expressed in terms of the simultaneous eigenstates of $C_{\pm}$, $L_z$ and $K_z$, 
\begin{equation}
\ket{(\ell_+,\ell_-); \omega, m} \equiv \ket{\ell_+,m_{+}}\otimes \ket{\ell_{-},m_{-}}\, ,
\end{equation}
such that
\begin{subequations}
\begin{align}
C_{\pm} \ket{(\ell_+,\ell_-); \omega,m} &= \ell_{\pm}(\ell_{\pm} + 1) \ket{(\ell_+,\ell_-); \omega, m}\,, \\[.5em]
A_z^{\pm} \ket{(\ell_+,\ell_-); \omega, m} &= m_{\pm} \; \ket{(\ell_+,\ell_-); \omega, m}\,, 
\end{align}
\end{subequations}
where $m_{\pm} = \frac{1}{2}(m \pm \ii \omega)$, with $m$ and $\omega$ being the eigenvalues of $L_z$ and $K_z$, respectively,
\begin{subequations}
\begin{align}
L_z \ket{(\ell_+,\ell_-); \omega,m} &= m \,\ket{(\ell_+,\ell_-); \omega,m}\, \\[.5em]
K_z \ket{(\ell_+,\ell_-); \omega,m} &= \omega \, \ket{(\ell_+,\ell_-); \omega,m}\,.
\end{align}
\end{subequations}

The quantum number $m$ takes half-integer values, while the eigenvalues of boosts $\omega$ take real values. From now on we will omit the representation labels $(\rho,k)$ and $(\ell_+,\ell_-)$ in the kets $\ket{j\, m}$ and $\ket{\omega \, m}$, respectively, to simplify the notation. The two bases are related by the \emph{overlap coefficients}, which were introduced by Husz\'ar \cite{Huszar:1971nn}
(see also \cite{Rashid:1979xv}), and take the following form
\begin{align}
	\bra{\omega\, m}\ket{j\,m} \equiv \bra{(\ell_+,\ell_-); \omega , m} \ket{(\rho,k);j,m} = 
    \ee^{-\ii \pi j} \, \ee^{+\ii \Psi_j^\rho}
    \sqrt{\frac{2j+1}{4\pi}} \; \mathfrak{U}_{jm}^{(\rho,k)} (\omega) \times 
	\begin{cases}
	1 & \text{if } k\leq m \\[.5em]
	\frac{\sin[\pi (\ell_{+}-m_{+})]}{\sin[\pi (\ell_{-}-m_{-})]} & \text{if } k>m 
	\end{cases}
	\label{eq:Overlap-Definition}
\end{align}
where the function $\mathfrak{U}_{jm}^{(\rho,k)}(\omega)$ is given by
\begin{align}
\mathfrak{U}_{jm}^{(\rho,k)} (\omega) &= \frac{ \Gamma(l'-m'+1) \; \Gamma(l'+m'+1)}{\Gamma(k'-\mu'+1) \; \Gamma(k'+\ii \rho+1)} \;  \nonumber \\
&\quad \times \sqrt{\frac{\Gamma(j+k'+1)\Gamma(j-\mu'+1)\Gamma(-\ell_{-}+\ell_{+}+j+1)  }{ \Gamma(j-k'+1)\Gamma(j+\mu'+1)\Gamma(\ell_{-}-\ell_{+}+j+1) }} \sqrt{\frac{\Gamma(\ell_{-}-m_{-}+1) \Gamma(\ell_{-}+m_{-}+1 )}{\Gamma(\ell_{+}-m_{+}+1)\Gamma(\ell_{+}+m_{+}+1) }} \nonumber \\
&\quad\times \; {}_3F_2[\{ -j+k',j+k'+1,l'-m'+1 \}, \{ k'-\mu'+1,k'+\ii \rho+1\}; 1] \, ,
\label{eq:Overlap-U}
\end{align}
and the following shifted variables have been introduced to simplify the notation \cite{Geiller:2014eza}:
\begin{subequations}
\begin{align}
k' &= \frac{1}{2}(\abs{k+m}+\abs{k-m}) \, , \quad \mu' = \frac{1}{2}(\abs{k+m}-\abs{k-m}) \, , \\
l' &= \frac{1}{2} (k'+\ii \rho - 1) \, , \qquad \quad \;\;\;m' = \frac{1}{2}(\mu'+ \ii \omega) \, .
\end{align}
\end{subequations}
The complete set of states $\ket{\omega\, m}$ is orthonormal, $\bra{\omega\,m}\ket{\omega'\, n} = \delta_{m n} \; \delta(\omega - \omega')$, and satisfy the resolution of the identity,
\begin{equation}
	\mathbbm{1} = \sum_m\int_{-\infty}^{\infty} \dd{\omega} \;  \ket{\omega\, m} \bra{\omega\,m} \, .
\end{equation}
By inserting the identity into the definition of the reduced Wigner $d$-matrix, we can write
\begin{align}
d_{jlm}^{(\rho,k)}(\beta) &= D^{(\rho,k)}_{jmlm}(\ee^{\tfrac{\beta \sigma_z}{2}}) \nonumber \\
&=\bra{j\,m}  \ee^{-\ii \beta \hat{K}_z}\ket{l\,m} \nonumber\\
&=\int_{-\infty}^{\infty} \dd{\omega}   \bra{j\,m} \ee^{-\ii \beta K_z} \ket{ \omega\, m} \bra{\omega\, m} \ket{l\,m} \nonumber \\
&= \int_{-\infty}^{\infty} \dd{\omega} \ee^{-\ii \beta \omega} \overline{\bra{ \omega\, m }\ket{j\, m}} \bra{ \omega\, m}\ket{l\, m} \, .
\label{eq:d-as-overlaps}
\end{align}
where the overlap coefficients follow R\"uhl's convention \eqref{eq:Overlap-Definition}; the conversion to the spinfoam-literature convention is given by $\Phi(\rho;j,l)$
 in \eqref{eq:other-phase}. From the first line in \eqref{eq:Overlap-U}, we note that the overlap coefficients $\bra{\omega\, m }\ket{j\,m}$ have simple poles in the complex $\omega$ plane. In particular, the integrand has four infinite series of poles located at 
\begin{align*}
\omega_1 &=  - \Big( \ii \, (1 + 2n+\abs{k-m}) -  \rho \Big)   \equiv \omega_n^{-} \\
\omega_2 &=  - \Big( \ii \, (1 + 2n+\abs{k+m}) + \rho \Big) \equiv \omega_n^{+}  \\
\omega_3 &=  \ii \,(1 + 2n+\abs{k+m}) -  \rho  \, \\
\omega_4 &=  \ii \,(1 + 2n+\abs{k-m}) +  \rho  \, .
\end{align*}
Since the asymptotic behavior of the integrand is $\ee^{-\ii \beta \omega} \ee^{-\pi \abs{\omega}}$ for $\abs{\omega} \to \infty$, we can close the integration contour in the lower-half complex $\omega$ plane for $\beta>0$. Here, the two relevant families of poles are given by $\omega_n^{-}$ and $\omega_n^{+}$. Integrating along the contour of integration covering the real axis, $\mathcal{C}$, which includes both families of poles, provides a representation for the reduced Wigner $d$-matrix, as shown in \eqref{eq:d-as-overlaps}. 

However, we can also deform the contour into two pieces, $\mathcal{C}_{+} \cup \mathcal{C}_{-} = \mathcal{C}$, such that $\mathcal{C}_{+} $ encloses the poles $\omega_n^{+}$ while  $\mathcal{C}_{-} $ encloses the poles $\omega_n^{-}$, as shown in Fig.~\ref{fig:Kz-Poles}. We find that to each contour integral there is a corresponding reduced Toller $t$-matrix:
\begin{equation}
t^{(\pm,\,\rho,k)}_{jlm} (\beta) = 
\int_{\mathcal{C}_{\pm}} \dd{\omega} \ee^{-\ii \beta \omega} \overline{\bra{ \omega\, m }\ket{j\, m}} \bra{ \omega\, m}\ket{l\, m} \, ,
\end{equation}
 Since we know the location of the poles, we can use Cauchy's residue theorem to compute the contour integrals, finding that 
\begin{align}
t^{(\pm,\,\rho,k)}_{jlm} (\beta) &=- 2\pi \ii \, 
\sum_{n=0}^{\infty} \ee^{-\ii \beta \omega_{n}^{\pm}}\, \Res[ \overline{\bra{\omega\, m} \ket{j\,m} } \bra{\omega\, m} \ket{l\,m} , \omega_n^{\pm} ] \, .
\end{align}
The same contour components \(t^{(\pm,\rho,k)}_{jlm}(\beta)\) admit the
coefficient-level representation derived in App.~\ref{app:Wick-rotation}
from the analytic continuation of the Euclidean Clebsch--Gordan expansion.

\section{Wick rotation and Clebsch-Gordan coefficients}
\label{app:Wick-rotation}

The self-dual decomposition of the complexified Lorentz algebra presented in \eqref{eq:self-dual-Lorentz} has a compact real form, defined by the real algebra $\mathfrak{so}(4) \cong \mathfrak{su}_{\mathbb{R}}(2) \bigoplus \mathfrak{su}_{\mathbb{R}}(2)$, which generates the group $Spin(4) \cong SU(2)\times SU(2)$. The algebra is generated by $\vec{J}_L$ and $\vec{J}_R$, such that
\begin{align}
[J_L^i, J_L^j] &= \ii \, \epsilon^{ij}_{~k} J^k_L \, , \nonumber \\
[J_R^i, J_R^j] &= \ii \, \epsilon^{ij}_{~k} J^k_R \, , \nonumber \\
[J_L^i, J_R^j] &= 0\, .
\end{align}
In this decomposition, the finite dimensional unitary irreducible representations are labeled by the pair of half-integers $(j_L,j_R)$, such that the two corresponding Casimir operators are
\begin{align}
C_L&=\vec{J}_L^{~2} = j_L(j_L+1)\, , \\
C_R &= \vec{J}_R^{~2} = j_R(j_R+1) \, .
\end{align}
Consequently, the associated Hilbert space admits the decomposition
\begin{equation}
    \mathcal{H}^{(j_L,j_R)} = \mathcal{H}_{j_L} \otimes \mathcal{H}_{j_R} \,,
\end{equation}
with $\dim(\mathcal{H}^{(j_L,j_R)}) = (2j_L+1)(2j_R+1)$. A decomposition which is directly analogous to  the Lorentzian case is also available, as we can write $\vec{J}_L=\tfrac{1}{2}(\vec{L}+\vec{M})$ and $\vec{J}_R=\tfrac{1}{2}(\vec{L}-\vec{M})$, where $\vec{L}$ is the generator of rotations that leave a fixed unit vector $t^I\in \mathbb{R}^4$ invariant, and $\vec{M}$ is the generator of rotations in the planes containing $t^I$, i.e., Euclidean boosts. With this identification, we can determine a basis in terms of the simultaneous eigenstates of $C_{L,R}$, $(J_{L,R})_z$, $L_z$ and $M_z$,
\begin{equation}
\ket{(j_L,j_R);\omega^{(E)},m} \equiv \ket{j_L,m_L} \otimes \ket{j_R,m_R}\, ,
\end{equation}
such that
\begin{subequations}
\label{eq:CanonicalBasis-jLjR-Euclidean}
\begin{align}
C_L\ket{(j_L,j_R);\omega^{(E)},m}&=j_L(j_L+1)\ket{(j_L,j_R);\omega^{(E)},m}\,,\\[.5em]
C_R\ket{(j_L,j_R);\omega^{(E)},m}&= j_R(j_R+1)\;\ket{(j_L,j_R);\omega^{(E)},m}\,,\\[.5em]
(J_L)_{z}\ket{(j_L,j_R);\omega^{(E)},m}&=m_L\ket{(j_L,j_R);\omega^{(E)},m}\,,\\[.5em]
(J_R)_{z}\ket{(j_L,j_R);\omega^{(E)},m}&= m_R \ket{(j_L,j_R);\omega^{(E)},m}\,,
\end{align}
\end{subequations}
where $m$ are the eigenvalues of $L_z$, and $\omega^{(E)}$ are the eigenvalues of the Euclidean boost along the $z$-axis, $M_z$,
\begin{align}
L_{z}\ket{(j_L,j_R);\omega^{(E)},m}&=m \ket{(j_L,j_R);\omega^{(E)},m}\,,\\[.5em]
M_{z}\ket{(j_L,j_R);\omega^{(E)},m}&= \omega^{(E)} \ket{(j_L,j_R);\omega^{(E)},m}\,,
\end{align}
such that $m_L = \tfrac{1}{2}(m+\omega^{(E)})$, and $m_R = \tfrac{1}{2}( m -\omega^{(E)})$. All these quantum numbers are half-integers, and this basis is the Euclidean equivalent of $\ket{(\ell_{+},\ell_{-});m,\omega}$ introduced above. To obtain a more explicit comparison with the Lorentzian version, we follow \cite{Dona:2021ldn} and introduce a new parametrization of the $\mathfrak{so}(4)$ algebra in terms of the generators $\vec{L}$ and $\vec{M}$, which satisfy the following commutation relations
\begin{align}
[L^i, L^j] &= \ii \, \epsilon^{ij}_{~k} L^k \, , \nonumber \\
[L^i, M^j] &= \ii \, \epsilon^{ij}_{~k} M^k \, , \nonumber \\
[M^i, M^j] &= \ii \, \epsilon^{ij}_{~k} L^k \, .
\end{align}
We identify two invariant operators, the Euclidean version of \eqref{eq:Casimirs-SL2C}, given by
\begin{subequations}
 \begin{align}
C_1^{(E)} &= -  \vec{M}^2-\vec{L}^2= -2(\vec{J}_L^{~2} + \vec{J}_R^{~2}) = -p^2 - q^2 + 1  \, , \\[0.5em] C_2^{(E)} &=\vec{M}\cdot \vec{L} =  \vec{J}_L^{~2} - \vec{J}_R^{~2} = p\, q \, ,
\end{align}
\end{subequations}
where we have introduced a new pair of half-integers, $(p,q)$, that characterize any given unitary representation, such that
\begin{equation}
j_L = \frac{1}{2} \qty(p+q-1) \, , \qquad j_R = \frac{1}{2} (p-q-1) \, .
\end{equation}
Using the labels $(p,q)$, the Hilbert space can be decomposed in $SU(2)$ subsectors with total angular momentum $\vec{L}^2 = (\vec{J}_L+\vec{J}_R)^2=j(j+1)$, such that
\begin{equation}
\mathcal{H}^{(p,q)} = \bigoplus_{j=\abs{q}}^{p-1} \mathcal{H}_j \, .
\end{equation}
In this representation space, we identify a canonical basis $\ket{(p,q);j,m}$ that simultaneously diagonalizes $\vec{L}^2$ and $L_z$, providing the Euclidean version of \eqref{eq:CanonicalBasis-jm}:
\begin{subequations}
\label{eq:CanonicalBasis-jm-Euclidean}
\begin{align}
C_1^{(E)}\ket{(p,q);j,m}&=(-p^2-q^2+1)\ket{(p,q);j,m}\,,\\[.5em]
C_2^{(E)}\ket{(p,q);j,m}&= p \, q\;\ket{(p,q);j,m}\,,\\[.5em]
\vec{L}^2\ket{(p,q);j,m}&=j(j+1)\ket{(p,q);j,m}\,,\\[.5em]
L_z\ket{(p,q);j,m}&=m \ket{(p,q);j,m}\,.
\end{align}
\end{subequations}
An explicit relation between these two bases is available, as we recognize that the Euclidean overlap coefficients are given by the standard $SU(2)$ Clebsch-Gordan coefficients,
\begin{align}
\label{eq:overlap-CG-Euclidean}
\bra{ \omega^{(E)}\, m}\ket{j\,m} &\equiv \bra{(j_L,j_R);\omega^{(E)},m } \ket{(p,q);j,m} \nonumber \\
&= C_{j_Lm_L\, j_R m_R}^{jm}\, .
\end{align}

The unitary representations of $h\in Spin(4)$ are described by a map $\mathcal{D}^{(p,q)}(h) : Spin(4)\to \mathcal{U}(\mathcal{H}^{(p,q)})$. In the magnetic basis, the elements of the Wigner $D$-matrix are given by
\begin{align}
D^{(p,q)}_{jm\, ln}(h) &= \bra{ (p,q);j,m } \mathcal{D}^{(p,q)}(h) \ket{ (p,q);l,n}\, .
\end{align}
It is useful to introduce a Cartan decomposition for an arbitrary element of the group $Spin(4)$, which can be completely characterized by
\begin{equation}
h = u_1 \, \ee^{\ii \, t \, M_z}\, u_2
\end{equation}
where $t \in [0,2\pi) $, and $u_1,\, u_2 \in \mathrm{diag} \, SU(2)$. This decomposition allows us to define the reduced Wigner $d$-matrices for $Spin(4)$, which after inserting a resolution of the identity, can be written as \cite{Biedenharn:1961drs,Barut:1976ub,Lorente:2003wn,Dona:2021ldn}
\begin{align}
&d_{jlm}^{(p,q)}(t) = D^{(p,q)}_{jmlm}(\ee^{\ii \, t \, M_z} ) \nonumber \\[0.5em]
&= \bra{(p,q);j,m} \ee^{\ii \, t M_z }  \ket{(p,q);l,m} \nonumber \\[0.5em]
&= \!\!\sum_{\omega^{(E)}=-j_L-j_R}^{j_L+j_R} \!\!\ee^{\ii\, t\,   \omega^{(E)}} \;\overline{\bra{ \omega^{(E)} \, m}\ket{j\,m}} \bra{ \omega^{(E)}\, m}\ket{l\, m} \nonumber\\
\label{eq:app-d-as-overlaps-Euclidean}
\end{align}

The Lorentzian overlap coefficients are intimately related to the analytic continuation of the $SU(2)$ Clebsch-Gordan coefficients to complex spins, which is defined by the Van der Waerden formula
\begin{align}
\mathfrak{C}(j_1 m_1,j_2 m_2 \,;\, j m)
&\nonumber \\
&\hspace{-8em}= \delta_{m,\, m_1 + m_2}\,
\sqrt{2j + 1}\,
\Biggl[
\frac{
\Gamma(j_1 + j_2 - j + 1)\,
\Gamma(j_1 - j_2 + j + 1)\,
\Gamma(-j_1 + j_2 + j + 1)
}{
\Gamma(j_1 + j_2 + j + 2)
}
\Biggr]^{\!1/2}
\nonumber\\[.5em]
&\hspace{-8em}\times
\Bigl[
\Gamma(j_1 + m_1 + 1)\,
\Gamma(j_1 - m_1 + 1)\,
\Gamma(j_2 + m_2 + 1)\,
\Gamma(j_2 - m_2 + 1)\,
\Gamma(j + m + 1)\,
\Gamma(j - m + 1)
\Bigr]^{\!1/2}
\nonumber\\[.5em]
&\hspace{-8em}\times
\sum_{t}
\frac{(-1)^{t} \Big(\Gamma(j - j_2 + m_1 + t + 1)\,
\Gamma(j - j_1 - m_2 + t + 1) \Big)^{-1}}{
t!\,
\Gamma(j_1 + j_2 - j - t + 1)\,
\Gamma(j_1 - m_1 - t + 1)\,
\Gamma(j_2 + m_2 - t + 1)\,
}\,,
\label{eq:VanDerWaerden-CG}
\end{align}
where the summation range is determined by the existence conditions of the Gamma functions in the denominator. For half-integer spins satisfying the usual $SU(2)$ admissibility conditions, the analytically continued coefficient reduces to the ordinary Clebsch-Gordan coefficient,
\begin{equation}
\mathfrak{C}(j_1,m_1;\,j_2,m_2;\,j,m) \rightarrow C^{jm}_{j_1m_1\,j_2m_2}\, .
\label{eq:VdW-reduction}
\end{equation}

\subsection{Coefficient-level continuation and compact-support thresholds}
\label{sec:thresholds}

The Wick rotation used in the main text is implemented on the two equivalent Euclidean Clebsch-Gordan expansions of the Biedenharn-Dolginov function as
\begin{align}
\label{eq:app-d-L}
d^{(p,q)}_{L\,jlm}(t)
&=
\sum_{n=0}^{2j_L}
\ee^{\ii t(2n-2j_L-m)}
\,
C^{jm}_{j_L(n-j_L)\,j_R(m-n+j_L)}
\,
C^{lm}_{j_L(n-j_L)\,j_R(m-n+j_L)}\, ,
\\[0.7em]
\label{eq:app-d-R}
d^{(p,q)}_{R\,jlm}(t)
&=
\sum_{r=0}^{2j_R}
\ee^{\ii t(m-2j_R+2r)}
\,
C^{jm}_{j_L(m-j_R+r)\,j_R(j_R-r)}
\,
C^{lm}_{j_L(m-j_R+r)\,j_R(j_R-r)}\, .
\end{align}
The two parametrizations $n=j_L+m_L$ and $r=j_R-m_R$ are identical in the Euclidean but become distinct after analytic continuation, and they are naturally adapted to the two Toller branches.

\paragraph*{Plus branch.}
Applying $\mathcal{W}_+$, the exponential factor in \eqref{eq:app-d-L} becomes
\begin{equation}
\ee^{\ii t (2n-2j_L-m)} \;\longrightarrow\;  \ee^{-\beta(2n+1-k-m-\ii \rho)}\, ,
\end{equation}
which decays for $\beta>0$, so the finite Euclidean sum extends to an infinite series over $n\in\mathbb{N}_0$. Finiteness of the factor 
\begin{equation}
\frac{1}{\Gamma(n-k-m+1)}\, ,
\end{equation}
in \eqref{eq:VanDerWaerden-CG} imposes the condition $n-k-m \le -1$. The first non-vanishing term therefore occurs at $n_+ \equiv \max(0,k+m)$, and writing $n=s+n_+$ with $s\in\mathbb{N}_0$, together with $2n_+ - k - m = \abs{k+m}$, the exponential becomes
\begin{equation}
\ee^{-\beta(1+2s+\abs{k+m}-\ii\rho)}\, .
\end{equation}

\paragraph*{Minus branch.}
Applying $\mathcal{W}_-$ to \eqref{eq:app-d-R}, the exponential factor becomes
\begin{equation}
\ee^{\ii t (m-2j_R+2r)} \;\longrightarrow\;  \ee^{-\beta(2r+1-k+m+\ii \rho)}\, .
\end{equation}
Finiteness of the factor
\begin{equation}
\frac{1}{\Gamma(r-k+m+1)}\, .
\end{equation}
then requires $n_- \equiv \max(0,k-m)$. Writing $r=s+n_-$ with $s\in\mathbb{N}_0$, together with $2n_- + m - k = \abs{k-m}$, the exponential becomes
\begin{equation}
\ee^{-\beta(1+2s+\abs{k-m}+\ii\rho)}\, .
\end{equation}

After including the $s$-independent R\"uhl-convention phases collected in the definition of $\mathfrak C^{(\pm,\rho,k)}_{jlm,s}$ in Eq.~\eqref{eq:Clebsch-Gordan-products-EtoL} of the main text, these two coefficient-level continuations reproduce the Toller series $t^{(\pm,\rho,k)}_{jlm}(\beta)$ as Eq.~\eqref{eq:t-as-Wick-rotation}. Stirling's formula shows that the analytically continued Clebsch-Gordan coefficients grow at most polynomially in $s$, so both series converge for every $\beta>0$.

The map between $Spin(4)$ and $SL(2,\mathbb{C})$ representations introduced in Sec.~\ref{sec:Wick-rotation} is summarized in Table~\ref{Tab:Wick-map}. Each Wick rotation $\mathcal W_{\pm}$ acts on the labels $(p,q,t)$ of the Euclidean group, on the self-dual spins $(j_L,j_R)$, and on the Casimir operators in a manner consistent with the Lorentzian principal series.

\begin{table*}[t]
	\caption{Map for the coefficient-level Wick rotation from Euclidean \(Spin(4)\) Wigner $d$ to Lorentzian \(SL(2,\CC)\) Toller matrices.}
	\label{Tab:Wick-map}
	\begin{ruledtabular}
		\begin{tabular}{c c c c}
			\rule{0pt}{3ex} Euclidean  & Lorentzian & $\mathcal W_+$ & $\mathcal W_-$ \\[.5em]
            \hline
            \rule{0pt}{4ex} \begin{tabular}{@{}c@{}}
$\vec{J}_L = \tfrac{1}{2}(\vec{L}+\vec{M})$ \\[0.3em]
$\vec{J}_R = \tfrac{1}{2}(\vec{L}-\vec{M})$
\end{tabular}  & $ \vec{A}^{\pm} = \tfrac{1}{2} (\vec{L}\pm \ii \vec{K}) $ & $\vec{M}\to \ii \vec{K}$ &$\vec{M} \to \ii \vec{K}$ \\[.7em]
            \hline
			\rule{0pt}{4ex} \begin{tabular}{@{}c@{}}
$C_1^{(E)}=-p^2-q^2+1$ \\[0.3em]
$C_2^{(E)}=p\,q$
\end{tabular} & \begin{tabular}{@{}c@{}}
$C_1=\rho^2-k^2+1$ \\[0.3em]
$C_2=\rho\, k$
\end{tabular} & \begin{tabular}{@{}c@{}}
$C_1^{(E)}\to C_1$ \\[0.3em]
$C_2^{(E)}\to \ii\, C_2$
\end{tabular} & \begin{tabular}{@{}c@{}}
$C_1^{(E)}\to C_1$ \\[0.3em]
$C_2^{(E)}\to \ii\, C_2$
\end{tabular} \\[.5em]
			\hline
			\rule{0pt}{4ex} $(p,q,t) \in (\tfrac{1}{2}\mathbb{Z},\tfrac{1}{2}\mathbb{Z},[0,2\pi))$ & $ (\rho,k,\beta) \in (\mathbb{R},\tfrac{1}{2}\mathbb{Z},\mathbb{R}^{+}) $ & $(p,q,t)\to  (\ii \, \rho,\,k,\,\ii\beta)$ & $(p,q,t)\to  (-\ii \, \rho,\,-k,\,\ii\beta)$ \\[.5em]
            \hline
            \rule{0pt}{4ex}$ j_L = \frac{1}{2} \qty(p+q-1) $ & $ \ell_{+} = \frac{1}{2}(k + \ii \rho -1)$ & $j_L\longrightarrow \ell_{+}$  &  $j_L\longrightarrow -1-\ell_{+}$ \\[.5em]
            \hline
            \rule{0pt}{4ex}$ j_R=\frac{1}{2} \qty(p-q-1)$ & $ \ell_{-}= \frac{1}{2}(k - \ii \rho -1)$ & $j_R\longrightarrow -1-\ell_{-}$ &$j_R\longrightarrow \ell_{-}$ \\[.5em]
            \hline
            \rule{0pt}{4ex} $d^{(p,q)}_{L\,jlm}(t)=d^{(p,q)}_{R\,jlm}(t)$ & $t^{(+,\rho,k)}_{jlm}(\beta)+t^{(-,\rho,k)}_{jlm}(\beta)=d^{(\rho,k)}_{jlm}(\beta)$ & $d^{(p,q)}_{L\,jlm}(t)\underset{\mathcal W_+}{\hookrightarrow} t^{(+,\rho,k)}_{jlm}(\beta)$ & $d^{(p,q)}_{R\,jlm}(t)\underset{\mathcal W_-}{\hookrightarrow} t^{(-,\rho,k)}_{jlm}(\beta)$
		\end{tabular}
	\end{ruledtabular}
\end{table*}

\subsection{Inverse Wick rotation}
\label{app:inverse-Wick-rotation}

The coefficient-level Wick rotation can be inverted, branch by branch. The inverse maps on the representation labels are
\begin{align}
\mathcal{W}_+^{-1}:\quad &\rho \to -\ii p\, ,\quad k \to q\, ,\quad \beta \to -\ii t\, ,
\label{eq:inverse-Wick-plus-map} \\[0.4em]
\mathcal{W}_-^{-1}:\quad &\rho \to +\ii p\, ,\quad k \to -q\, ,\quad \beta \to -\ii t\, .
\label{eq:inverse-Wick-minus-map}
\end{align}
Under these maps, the Lorentzian infinite sums truncate by the compact
\(SU(2)\) support conditions. Before removing the convention factors, one finds
\begin{align}
t^{(\pm,\rho,k)}_{jlm}(\beta) \underset{\mathcal W_\pm^{-1}}{\hookrightarrow} \begin{cases}
\Pi_{jl}^{(-\ii p,q)}
\,d^{(p,q)}_{L\,jlm}(t),
\\[0.5em]
(-1)^{j-l}\Pi_{jl}^{(+\ii p,-q)}
\,d^{(p,q)}_{R\,jlm}(t).
\end{cases}
\end{align}
The phase convention matching R\"uhl's conventions is
\begin{equation}
\Pi_{jl}^{(\rho,k)}
\equiv
(-1)^{2k}
\ee^{\frac{\ii\pi}{2}(j-l)}
\ee^{-\ii\Psi_j^\rho+\ii\Psi_l^\rho}.
\end{equation}
This is the total \(s\)-independent phase multiplying the analytically
continued Clebsch--Gordan product in
Eq.~\eqref{eq:Clebsch-Gordan-products-EtoL}.  The $(-)$ branch also
contains the external factor \((-1)^{j-l}\) as in
Eq.~\eqref{eq:t-as-Wick-rotation}.

For the $(+)$ branch, the Lorentzian threshold
$n_+=\max(0,k+m)$ becomes $n_+^E=\max(0,q+m)$,
and the inverse is
\begin{align}
\left(\Pi_{jl}^{(\rho,k)}\right)^{-1}
t^{(+,\rho,k)}_{jlm}(\beta)
&\underset{\mathcal W_+^{-1}}{\hookrightarrow}
\sum_{s=0}^{2j_L-n_+^E}
\ee^{\ii t(2(s+n_+^E)-2j_L-m)}
\nonumber\\
&\qquad\times
C^{jm}_{j_L(s+n_+^E-j_L)\,j_R(m-s-n_+^E+j_L)}
C^{lm}_{j_L(s+n_+^E-j_L)\,j_R(m-s-n_+^E+j_L)}
\nonumber\\
&=
d^{(p,q)}_{L\,jlm}(t).
\end{align}
For the $(-)$ branch, the Lorentzian threshold
\(n_-=\max(0,k-m)\) becomes $n_-^E=\max(0,-q-m)$, and
\begin{align}
\left((-1)^{j-l}\Pi_{jl}^{(\rho,k)}\right)^{-1}
t^{(-,\rho,k)}_{jlm}(\beta)
&\underset{\mathcal W_-^{-1}}{\hookrightarrow}
\sum_{s=0}^{2j_R-n_-^E}
\ee^{\ii t(m-2j_R+2(s+n_-^E))}
\nonumber\\
&\qquad\times
C^{jm}_{j_L(m-j_R+s+n_-^E)\,j_R(j_R-s-n_-^E)}
C^{lm}_{j_L(m-j_R+s+n_-^E)\,j_R(j_R-s-n_-^E)}
\nonumber\\
&=
d^{(p,q)}_{R\,jlm}(t).
\end{align}
Since \(d^{(p,q)}_{L\,jlm}(t)=d^{(p,q)}_{R\,jlm}(t)\), the two inverse continuations recover the same Wigner $d$-matrix.

\vfill
\twocolumngrid


\providecommand{\href}[2]{#2}\begingroup\raggedright\endgroup


\end{document}